# Single microtubules and small networks become significantly stiffer on short time-scales upon mechanical stimulation

Matthias D. Koch[a,1], Natalie Schneider[b], Peter Nick[b], and Alexander Rohrbach[a,*]

[a] Laboratory for Bio- and Nano-Photonics, Department of Microsystems Engineering, University of Freiburg, Georges-Koehler-Allee 102, 79110 Freiburg, Germany

[b] Molecular Cell Biology, Botanical Institute, Karlsruhe Institute of Technology, Kaiserstr. 2, 76131 Karlsruhe, Germany

[1] Present address: Lewis-Sigler Institute for Integrative Genomics, Princeton University, Washington Rd, Princeton, NJ 08544, USA

[*] Correspondence: rohrbach@imtek.de, phone: 0049 761 203 7536

**Abstract**

The transfer of mechanical signals through cells is a complex phenomenon. To uncover a new mechanotransduction pathway, we study the frequency-dependent transport of mechanical stimuli by single microtubules and small networks in a bottom-up approach using optically trapped beads as anchor points. We interconnected microtubules to linear and triangular geometries to perform micro-rheology by defined oscillations of the beads relative to each other. We found a substantial stiffening of single filaments above a characteristic transition frequency of 1-30 Hz depending on the filament's molecular composition. Below this frequency, filament elasticity only depends on its contour and persistence length. Interestingly, this elastic behavior is transferable to small networks, where we found the surprising effect that linear two filament connections act as transistor-like, angle dependent momentum filters, whereas triangular networks act as stabilizing elements. These observations implicate that cells can tune mechanical signals by temporal and spatial filtering stronger and more flexibly than expected.

## Introduction

Today, we know that cells across all domains are mechanosensitive [1], and that mechanosensitivity is the base for sensing quite different stimulus qualities including osmotic challenges, gravity, movements or even sound. In addition, mechanosensitivity is used to organize and integrate cells and organs into functional units, e.g., in the course of movements in metazoan organisms or during plant development [2]. Perturbations of mechanotransduction have been implicated in various severe diseases like cancer [3, 4]. Remodeling of the cell as a response or adaption to an external, physical stimulus is steered by gene expression in the nucleus [5]. Therefore, the information of the stimulus has to be transported across the cell from the periphery to the center. Common models of cellular mechanotransduction assume the conversion of a physical stimulus to a chemical signal by membrane proteins such as integrins [3], and the subsequent transport to the nucleus either passively by diffusion or actively by molecular motors, i.e., rather slow processes. However, the direct propagation of a mechanical stimulus by stress waves through stiff cytoskeletal elements connecting the membrane and the nucleus [6] would enable a much faster transport pathway on the microsecond timescale and thus allow almost instantaneous integration of responses across the cell [7]. A model for such a pathway has been proposed by Ingber [8, 9] on the basis of a tensegrity model of flexible actin filaments (able to transmit traction forces) connected to the relatively stiff microtubules (able to transmit compression forces). In mammalian cells, microtubules are typically aligned radially inside a cell spanning from the centrosome, located close to the nucleus, to the cell membrane [10], a set up that would allow for efficient mechanotransduction between cell membrane and nucleus [11]. In fact, mechanical stimulation has been shown recently to induce a perinuclear actin ring, brought about by the activity of actin-microtubule cross-linking formins [12]. MTs are well known as components of mechanosensing in flies [13] as well as in vertebrates [14] and microtubules have also been found to participate in gravity sensing and mechanic integration in plants (reviewed in [15]). Remarkably, mutants of *Caenorhabditis* affected in beta tubulin turned out to be insensitive to mechanic stimulation [16]. Efficiency and specificity of the MT sensory functions, however, depend on their frequency dependent viscoelastic properties, which are characteristic for biological systems.

To address these aspects of microtubule-dependent signaling, we present an approach for the targeted construction of cytoskeletal meshes with defined geometries by using optically trapped beads as anchor points. Existing approaches only demonstrated the construction of small networks without biologically relevant measurements[17] or rely on the stochastic attachment or growth of filaments to optically trapped beads or micro pillars which is less flexible and barely allows control of the number of attached filaments [18, 19, 20]. We use established micro-rheology techniques [21, 22, 23] to measure the time-dependent viscoelastic properties of single microtubules. Existing approaches have investigated bulk material properties [24, 25, 26] *in vitro*, the properties of the cytoskeletal molecules *in vivo* [27] ignoring their organization or number [28, 29, 30], or static representations of single filaments [31, 32]. Microtubule stiffness and bending relaxation has also been addressed by Fourier decomposition of bending modes caused by thermal fluctuation [33, 34], however, thermal forces are not sufficient in amplitude to significantly deform a whole network. A recent review summarizing the molecular origins of microtubule mechanics and highlighting effects of network architecture during stress transmission is presented by Lopez and Valentine[35].

Since different cell types show different structures of the cytoskeleton and alignment of microtubules, we test the viscoelastic properties of different small network topologies on their performance of conducting mechanical stimuli at different frequencies. This approach should not only allow determining which network symmetries are best suited to transduce mechanical signals, but also to get insight into the general role of the cytoskeletal structure and function in different cell types by successively increasing the complexity of the network through addition of other cytoskeletal associated components.

**Results**

To determine the time-dependent viscoelastic properties of single microtubules (MTs) and small networks of MTs, movable Neutravidin coated beads as anchor points were attached to a biotinylated microtubule at defined positions by time-shared optical tweezers (see Methods). Then, these anchor points were mutually displaced in an oscillatory fashion with defined frequencies and amplitudes along the x-direction as illustrated in Fig. 1. The resulting frequency dependent stretching and buckling behavior of these constructs is measured, which allows determining both the elastic and the viscous properties of the MT constructs in different geometrical arrangements.

**Stiffening of single filaments at high oscillation frequencies**

Upon force generation, the beads are displaced from their equilibrium position with a straight microtubule as depicted in Fig. 1. The displacements $x_{B1}(t) – x_{L1}(t)$ and $x_{B2}(t) – x_{L2}(t)$ of bead 1 (actor) and bead 2 (sensor) relative to the laser trap positions $x_{L1}$ and $x_{L2}$, are shown exemplarily in Fig. 2a,b for two different actor displacement frequencies, $f_a = 0.1$ Hz and $f_a =$ 100 Hz, at a displacement amplitude $A_a$ = 500 nm. Due to high tensile and small buckling forces, the sensor bead is pulled out of the trap center by up to $x_{B2}$ ≈ 80nm and pushed only slightly by less than $x_{B2}$ ≈ 10 nm during each half period. This situation changes significantly at high frequencies $f_a$ = 100Hz. While the maximum displacements $x_{B2}$ during microtubule stretching were approximately the same at $f_a$ = 100 Hz and $f_a$ = 0.1 Hz, the displacement increased by an order of magnitude at high frequencies during buckling, i.e., $x_{B2}(f_a = 100 \text{ Hz})$ ≈ 10·$x_{B2}(f_a = 0.1 \text{ Hz})$. Therefore, only the compression and buckling of single filaments will be analyzed in this study. The complete frequency dependence of the filament – bead construct is expressed by the average maximum distance change between both beads $\Delta L_x = A_a - x_{max1} - x_{max2}$ as shown in Fig. 2c for three different amplitudes $A_a$ during stretching and buckling. While the slight amplitude decrease during stretching can be attributed to the increasing friction force $F_{\gamma,B} = 6\pi R \eta \dot{x}_B$ of the bead, the much stronger drop during buckling is caused by the microtubule filament, indicating an apparent stiffening of the filament at high oscillation frequencies $f_a$ > 1 Hz (see Supplementary Results Fig. S2). We would like to point out that microtubules are inextensible and not actually stretched during the pulling phase [32]. Instead, they are bent locally as shown in Fig. 1a. To estimate whether this frequency response is a purely viscous effect governed by the friction of the actuated filament, we analyzed the amplitudes of the involved forces theoretically as explained in the Supplementary Results and shown in Fig. 2d. The main contributions to the amplitude of the total viscous force $F_{\gamma,tot} \approx F_{\gamma_B,tran} + F_{\gamma_{MT},\perp}$ are the translational viscous forces of the beads $F_{\gamma_B,tran}$ and the perpendicular viscous force component of the filament $F_{\gamma_{MT},\perp}$. Both, the rotational viscous forces of the beads $F_{\gamma_B,rot}$, acting as hinged supports of the filament ends, and the parallel viscous forces $F_{\gamma_{MT},\parallel}$, are negligible. A comparison between the theoretically estimated viscous and the experimentally obtained total force amplitudes reveals a strong difference especially dominant at intermediate frequencies 1 Hz ≤ $f_a$ ≤ 100 Hz. This shows that significant elastic forces control the deformation of the microtubule filament.

**Excitation and relaxation of higher MT deformation modes**

As introduced above, the oscillatory driving force counteracts against the viscous and the elastic forces of both the MT and the two beads. The behavior of the semi-flexible MT of length $L$ is described by the hydrodynamic beam equation, which predicts that induced MT deformations can be described by a superposition of sine waves with wave numbers $q_n = \frac{n \cdot \pi}{L}$ and a characteristic relaxation time proportional to $1/q^4 \sim L^4$ (see Methods and Supplementary Results). Hence, higher deformation modes n > 1 can only be excited at higher driving frequencies $\omega = 2\pi \cdot f_a$, leading to the effect of MT stiffening. The stiffening can be described by the frequency dependent complex shear modulus $G(\omega) = G'(\omega) + i \cdot G''(\omega)$ (see Methods Section), where a representation of all forces in frequency space allows to extract the elastic component $G'(\omega)$ and the viscous component $G''(\omega)$.

The elastic modulus $G'(\omega)$ shown in Fig. 3 describes the frequency dependent MT stiffness, which is characterized by a constant plateau value $G'(0)$ and a frequency dependent response $G'(\omega \gg \omega_1)$ at high frequencies. This can be estimated as

$$G'(\omega = 0) = \tfrac{\pi^2}{2.16} \cdot \left(\tfrac{1}{L}\right)^4 \ell_p k_B T \quad \text{and} \quad G'(\omega > \omega_1) \sim \omega^p \qquad (1)$$

Here, $l_p = EI / k_B T$ is the persistence length of a semiflexible polymer with $L \ll l_p$ and bending modulus $EI$ (flexural rigidity). The frequency of the MT's ground mode $\omega_1 = \tfrac{EI}{g_{MT}} \left(\tfrac{\pi}{L}\right)^4$ depends on the viscous drag $g_{MT}$ of the MT. As shown further below, we obtain a power law exponent $p = 5/4$ for oscillations of single microtubules in longitudinal direction, matching the theoretical prediction for semiflexible filaments in the intermediate frequency regime [36, 37] tested here. However, we also find that the degree of stiffening, the exponent $p$, depends on the filament stabilization, i.e., the molecular architecture of the filament, and the direction of oscillation.

We checked whether the measured frequency response and apparent stiffening of single filaments indeed results from the excitation of higher deformation modes as described by equation (1). Therefore, we analyzed the dynamics of the trapped anchor points with two particle active micro-rheology techniques (see [22, 23] and Supplementary Methods for details) as described in the following.

The frequency dependent elastic response of the single microtubule was analyzed in terms of $G'(\omega)$. As explained in the Supplementary Information, we can assume proper linear response for all of the mentioned experimental conditions. The results for filaments stabilized with Taxol and polymerized with either GTP or its slowly hydrolysable homolog GMPCPP are displayed in Fig. 3a-c. Here, the results are grouped according to lengths $L \approx 5\mu m$ (3.4μm ≤ $L$ ≤ 6.0μm for Taxol) and $L \approx 15$ μm (15μm ≤ $L$ ≤ 25μm for Taxol and 10μm ≤ $L$ ≤ 16μm for GMPCPP). Each group represents the average of 3-6 individual filaments each probed at 2-4 different oscillation amplitudes ($A_a$ = 200nm, 400nm, 600nm typically), resulting in approximately 10 measurements per group. The grouping was chosen due to the length dependence of the persistence length [34, 38], as discussed further below, and because no significant difference in elasticity within these individual groups could be observed. For an oscillation parallel to the filament axis, the theoretical slope with $p = 1.25$ according to equation (1) (also see Methods) fits well to our experimental results as shown in Fig. 3a-c. However, we also used a free exponent $p$ as additional fit parameter, $G'(\omega) - G'(0) = C\omega^p$, to check for any deviation from the theoretical prediction. We obtained $p = 1.08 \pm 0.15$ and $p = 0.93 \pm 0.07$ for short and long MTs stabilized by 10 µM Taxol, $p = 0.78 \pm 0.25$ and $p = 1.46 \pm 0.13$ for short and long MTs stabilized by 100 µM Taxol, and $p = 1.51 \pm 0.13$ for (long) GMPCPP filaments. Within the error margins this indicates a rough coincidence with our model (p = 0.75) for short MTs, whereas the stiffening for longer MTs better matches with the advanced model [36, 37] predicting an exponent $p = 1.25$.

**Frequency response depends on MT stabilizations**

**Frequency dependent persistence length.** Fig. 3a-c show that the plateau value depends on the length of the filament and filament stabilization. For long filaments ($L \approx 15$µm) we find $G'(0) = 1.5$ mPa (with 10 µM Taxol), $G'(0) = 2$ mPa (with 100µM Taxol) and $G'(0) = 4$ mPa (with GMPCPP). For short filaments ($L \approx 5$ µm), MT stabilization hardly affects the plateau values, which are $G'(0) \approx 10$ mPa for both 10 µM and 100 µM Taxol. No data is available for short GMPCPP filaments. As shown in equation (1), the plateau value is a measure of the persistence length. Dividing $G'(\omega)$ by $\frac{\pi^2 kT}{2.16 \cdot L^4}$, we find a frequency dependent persistence length

$$\ell_p(\omega) = G'(\omega) \cdot \frac{2.16}{kT\pi^2} L^4 \approx \ell_p(0) + \ell_p(\omega > \omega_1), \qquad (2)$$

which increases with frequency because of a successive excitation of higher modes at $\omega > \omega_1$. It can be seen in Fig. 3d that the persistence length depends sensitively on the contour length $L$ of the MT. We find $l_p(0) = (0.33 \pm 0.05)$ mm for $L = 5$ µm and all stabilizations. For $L = 15$ µm we find $l_p(0) = (4.06 \pm 0.26)$ mm stabilized with 10 µM Taxol, $l_p(0) = (5.80 \pm 0.39)$ mm for 100 µM Taxol, and $l_p(0) = (12.10 \pm 0.66)$ mm for GMPCPP. The estimates based on equation (2) agree well with the published dependency of $l_p$ on the filament contour length [38] and stabilization [39], as further elucidated in the discussion.

**Transition frequency.** Beyond a characteristic frequency, a visible increase of $G'(\omega)$ is manifested due to the excitation of higher deformation modes. We define this transition by the frequency $\omega_t$ where $G'(\omega_t) / G'(0) = 1.5$, i.e., $G'(\omega)$ is increased by 50% such that

$$\omega_t \approx 3\omega_1 = \frac{3}{g_{MT}}\left(\frac{\pi}{L}\right)^4 \ell_p(0) kT \qquad (3)$$

We find that the transition frequency $\omega_t = 2\pi f_t$ scales by a factor of 3 with the ground mode. We obtain $f_t = (5.3 \pm 1.2)$ Hz and $f_t = (4.2 \pm 1.7)$ Hz for short filaments ($L = 5\mu m$) stabilized with 10 µM or 100 µM Taxol, respectively. For long filaments ($L = 15\mu m$), we find $f_t = (0.6 \pm 0.1)$ Hz, $f_t = (1.8 \pm 0.4)$ Hz and $f_t = (3.9 \pm 0.8)$ Hz for stabilization with 10 µM, or 100 µM Taxol, or for 100 µM Taxol + GMPCPP, respectively. Comparing this to the theoretical estimate of eq.(3) predicting $\omega_t = 21$ Hz and $\omega_t = 35$ Hz for short MTs with 10 µM and 100 µM and $\omega_t = 2$ Hz, $\omega_t = 3.3$ Hz and $\omega_t = 7$ Hz for long MTs with 10µM, 100 µM and GMPCPP, out experimentally obtained values are throughout too small, but follow the general dependence on length and stiffness (see Supplementary Results Fig. S11). Thus, the transition frequency $\omega_t$ increases with increasing stabilization, clearly indicating the importance of the molecular structure of microtubules (see Discussion). As indicated in Fig. 3a-c (vertical dashed lines), these results fit well to a graphical estimation of $f_t$ at $G'(\omega_t) = 1.5 \cdot G'(0)$.

**Lateral oscillation of single filaments.** So far, only longitudinal oscillations of filament ends have been considered, i.e. bead displacements parallel to the filament axis. Fig. 3c further shows the result for a bead oscillation lateral to the axes of long ($L = 15\mu m$) GMPCPP stabilized filaments. Again, we observe a plateau value for frequencies $\omega < \omega_t$ and a power law rise for $\omega > \omega_t$ with $p = 1.76 \pm 0.02$, i.e., a 40% larger stiffening exponent than the predicted value $p = 1.25$ for a longitudinal oscillation. In contrast, the plateau $G'_\perp(0) = 0.54$ mPa is approximately one order of magnitude smaller than $G_{//}'(0)$ in axial direction. The transition frequency $f_t = 3$ Hz obtained from $G'_\perp(\omega_t) = 1.5 G'_\perp(0)$ is approximately the same as in axial direction.

**Momentum transport along a linear chain of connected MTs**

An important question is whether the findings for single filaments can be used to predict the momentum transport through small networks of filaments - in analogy to Kirchhoff's circuit laws for the connection of currents in network nodes. However, for connected microtubules, i.e. for different networks, the compression of one filament usually results in a stretching of another filament and vice versa, such that a separation of compression and stretching is not possible anymore. Therefore, the complete oscillation period of the actor and sensor beads will be analyzed in the following.

In a first step, we constructed a linear network consisting of three optically trapped beads and two microtubule filaments as shown in Fig. 4A. This construct was probed such that trap 1 was oscillated sinusoidally at varying frequency and amplitude, while trap 2 and 3 remained stationary. In this way, we investigated the momentum transfer along the first microtubule, while attached to a second microtubule, through $G'^{(1,2)}(\omega)$, but also the momentum transfer along both microtubules through $G'^{(1,3)}(\omega)$. The longitudinal and lateral oscillation of the actor and sensor beads displayed in Fig. 4b,c reveal a qualitatively similar momentum transfer as for single filaments. Results are the average of three measurements for filaments of length $L = 10$ µm after stabilization with 100 µM Taxol.

**Longitudinal chain oscillation:** By inspecting the curves in Fig. 4b,c, both a frequency-independent plateau $G'^{(i,j)}(0)$, and a frequency-dependent behavior $G'^{(i,j)}(\omega)$ between beads $i$ and $j$ can be observed. Using power law fits according to equation (1), the plateau values $G'^{(1,2)}(0) = 6$ mPa and $G'^{(1,3)}(0) = 32$ mPa for a single (1→2) and two-step (1→3) MT connection are determined. Interestingly, the static elasticity $G'^{(1,3)}(0) > G'^{(1,2)}(0)$ for the two step connection is larger than for the single step. As shown in Fig. 4d, $G'^{(1,3)}(0, \kappa_{T2})$ is a function of the stiffness $\kappa_{T2}$ of the intermediate trap 2. Here, we repeated the measurement and doubled the trap stiffness $\kappa_{T2}$ each time, resulting in an increase of $G'^{(1,3)}(0) \sim \sqrt{\Delta \kappa_{T2}}$ proportional to the square root of the stiffness change $\Delta \kappa_{T2}$. While $G'^{(1,2)}(0)$ for the first 10 µm long filament fits well between $G'(0)$ for $L = 5$µm and $L = 15$µm shown above, $G'^{(1,3)}(0)$ for two 10 µm long filaments is much larger than $G'(0)$ for a single filament with $L = 20$ µm The transition frequency $f_t^{(1,2)} = (4.1 \pm 0.3)$ Hz (black dashed line in Fig. 4b) according to $f_t = \frac{1}{2g_{MT}} \left(\frac{\pi}{L}\right)^4 \ell_p(0) kT$ in equation (3) is approximately identical to that of single filaments, while $f_t^{(1,3)} = (17 \pm 2.5)$ Hz is approximately 4 times larger.

**Lateral chain oscillation:** As shown in Fig. 4b,c, the static elasticity $G'(0)$ for lateral displacements of beads and MT ends is significantly different to longitudinal (parallel) displacements. For the longitudinal elasticity $G_\parallel'^{(i,j)}(0)$, the double MT connection 1→3 was about five times stiffer than the direct MT connection 1→2, whereas for the lateral elasticity $G_\perp'^{(i,j)}(0)$, the connection 1→3 is about five times softer than the connection 1→2, i.e., $G_\perp'^{(1,3)}(0) \approx \frac{1}{5} G_\perp'^{(1,2)}(0)$. Beyond the transition frequency, the frequency dependent elasticity $G_\perp'^{(i,j)}(\omega) \sim \omega^p$ increases according to a power law exponent $p = 2.5 \pm 0.1$ and $p = 2.4 \pm 0.1$ for the connection 1→2 and 1→3. Interestingly, the transition frequency $f_t = (11 \pm 1)$ Hz is approximately the same for both connections, in contrast to the longitudinal oscillations. However, during a lateral oscillation, both filaments are always slightly stretched compared to a longitudinal oscillation, where filaments are buckled. The role of the intermediate connection and the role of the coupling point (trapped bead #2) are explained in the discussion.

**Momentum transport in an equilateral triangle**

We used GMPCPP filaments to construct equilateral triangles of 15 µm side length as depicted in Fig. 5a. The trap 1 is again oscillated in *x* or *y*, resulting in a trap movement radial or tangential to the connection between bead 1 and the center of the triangle. An overlay of brightfield and fluorescence images of one radial oscillation period at $f_a = 0.1$ Hz ($T = 1 / f_a = 10$ s) and $A_a = 600$ nm along *x* is shown in Fig. 5b. In contrast to single filaments and the linear chain, here, in total two filaments are always buckled or tense, while at the same time, the third one behaves in the opposite manner, i.e., is tense or buckled.

**Triangles are stiffer than single filaments and have a similar high frequency response**

Due to the symmetric configuration of the equilateral triangle, the elastic modulus for both connections 1→2 and 1→3 should be identical, except for different oscillation directions. This is indeed the case as shown in Fig. 6a,b for an exemplary construct, where the radial and tangential elastic responses, $G_x'^{(i,j)}(\omega)$ and $G_y'^{(i,j)}(\omega)$, are plotted for an actor bead oscillation along *x* and *y*. The static elasticities $G_x'^{(1,2)}(0) \approx G_x'^{(1,3)}(0) = (55 \pm 20)$ mPa and $G_y'^{(1,2)} \approx G_y'^{(1,3)} = (30 \pm 10)$ mPa indicate a 25-fold increase of the overall stiffness of the construct compared to that of single filaments (see to Fig. 3c). In the Supplementary Results, we present further data of triangular constructs with slight pretension, induced by thermal fluctuations of the filaments during construction of the network.

According to equation (3), the larger static elasticities should result in an increase of the transition frequency $\omega_t$ by a factor 25, i.e., $f_t = (810 \pm 308)$ Hz and $f_t = (75 \pm 13)$ Hz for both oscillation directions. 810 Hz is much larger than the measured maximum frequency, so that we cannot observe a power law rise for an oscillation along $x$. However, the extrapolated intersection of the single filament response (fit with free exponent according to equation (1)) with the plateau of the triangle can be estimated to $f_t \approx 800$ Hz, which is in good agreement with the theoretical estimate of equation (3). For the tangential oscillation direction ($y$) displayed in Fig. 6b, the network stiffens already at a transition frequency $f_t \approx (100 \pm 10)$ Hz (vertical black dashed line). This value is slightly larger than predicted (vertical red dashed line), which could be a consequence of the dip in $G'(\omega)$ at $f = 50$ Hz and might influence the value of $f_t$. This dip is not visible for other constructs with the same geometry as presented in the Supplementary Results.

## Discussion

**Microtubule stiffness depends on the contour length.** We have analyzed the elastic behavior of single and inter-connected MTs by means of the elastic modulus $G'(\omega)$, which can be described by a low frequency plateau $G'(0) \sim \ell_p(0)$, and a rise at high frequencies above a characteristic transition frequency $\omega_t$, defined by a 50% increase of $G'(\omega)$. Varying the molecular composition of the filaments, by stabilization agents had no visible effect on the static elasticity of short MTs (5 µm length). Interestingly, this was different for long MTs (of around 15µm length), where the effect of chemical stabilization on elasticity became visible, leading to the conclusion that the molecular coupling length extends over several µm. Similar effects have already been reported by Pampaloni et al.[38], who found a length dependence of the persistence length, which levels to a plateau above a critical length $\ell_c = 21$µm, i.e., $\ell_p(L > \ell_c)$ = const. Similarly, Taute et al.[40] introduced an additional internal friction term to explain deviations of their measured MT drag coefficients when microtubules were shorter than ≈ 5 µm, attributed to dissipation during conformational changes or liquid flow passing through narrow pores in the MT lattice as introduced by [41]. Polymorphic conformational states of the tubulin lattice and non-equilibrium filaments dynamics have also been studied in motor based microtubule gliding assays, recently[42, 43]. Irrespective of the effects found here, the plateau value $G'(\omega = 0)$ is directly related to the conventional, frequency independent persistence length $l_p(\omega = 0)$, which increases with the contour length of the MTs according to $l_p(\omega = 0, L) \sim 1 / (1 + l_c^2 / L^2)$ [38]. We measured two different ranges of lengths for single MTs varying in length by a factor of 3 (plus one intermediate length for the linear network). Considering the total length dependence $G'(0) = \frac{kT\pi^2}{2.16} \ell_p(0,L)\left(\frac{\pi}{L}\right)^4 \sim \left(\frac{1}{L}\right)^2$, which is approximately quadratic and results in a 9-fold higher plateau for 3-fold shorter MTs, we find a reasonable match with our measurements shown in Fig. 3. From the two MT lengths, we also find that our results for $\ell_p(0, L)$ agree well to those reported previously[34, 38].

**Frequency dependent persistence length and stiffness.** The novelty of our observations is the increase of the persistence length, or correspondingly the elastic modulus $G'(\omega)$, of a single microtubule with the displacement frequency $\omega$ (Fig. 3). In the Methods section, we show that this is caused by the excitation of higher deformation modes, which means that filaments become stiffer on shorter timescales, such that filament buckling is suppressed. In other words, molecular relaxation processes as a consequence of internal stress along the MT cannot follow on too short timescales. The timescale of molecular relaxation is approximated by the transition frequency $\omega_t \approx 3 \cdot \omega_{n=1}$, which we indicated in all plots of $G'(\omega)$. Beyond this frequency, the second deformation mode ($n = 2$) renders the filament about 1-4 times stiffer, beyond $\omega = 20 \cdot \omega_{n=1}$ the third deformation mode (n = 3) stiffens the filament 4-10 times relative to $\omega = 0$ as explained in the Supplementary Results (Fig. S4). Our measurements confirm the general, theoretically predicted trend of a smaller transition frequency for longer MTs. According to $\omega_t = \frac{3kT}{g_{MT}} \ell_p(0) \left(\frac{\pi}{L}\right)^4 \sim \left(\frac{1}{L}\right)^2$, we expect an about 9-fold lower transition frequency for 3-fold longer MTs (see Methods), which we obtain for 10μM Taxol stabilization, but not for 100μM Taxol, indicating that further theoretical studies are necessary.

For short MTs we measured a stiffness increase according to $G'(\omega) \sim \omega^{3/4}$ at high frequencies (50 Hz < $\omega$ < 100 Hz), whereas for longer MTs we found $G'(\omega >> \omega_t) \sim \omega^{5/4}$, which is reasonably close to the theoretical prediction based on the hydrodynamic beam model for semiflexible filaments [36, 37]. In some cases, though, such as for the strong GMPCPP stabilization, the fitting of a free exponent describes a power law behavior of $\omega^{1.5}$, which represents faster stiffening, i.e., a slower molecular relaxation on shorter timescales. It is known that MTs polymerized in the presence of slowly or non-hydrolyzable GTP analogs such as GMPCPP or γ-S-GTP form more lateral inter-protofilament contacts between β-tubulins as compared to GTP/GDP MTs (see [44, 45, 46] and the Supplementary Material). Approximating the connection between individual αβ- tubulin dimers by damped harmonic springs [47, 48], the damping of the intermolecular connections should affect the temporal response to mechanical stimuli and thereby the transition frequency $\omega_t$.

Within the accuracy of our measurements, the viscous modulus $G''(\omega)$ (see Supplementary Results) increases linearly with frequency as predicted by the lateral friction coefficient $g_{MT} \cdot \omega = \frac{4\pi\eta\omega}{\ln(L/D)+0.84}$ of a simple rod moving in an aqueous solution. Hence, we can clearly exclude that the strong increase of filament stiffness $G'$ at high frequencies is governed by simple friction on the filament.

In biological and other noisy systems, the signal energy stored in various degrees of freedom (translation, oscillation, etc.) is significantly less pronounced at higher frequencies (e.g., a $1/\omega^2$ decay for thermal motion). In this way, microtubules should act as transmission amplifiers or high pass filters for mechanical signals, based on our observations that mechanical stimuli are transferred much more efficiently at higher frequencies.

**Angular momentum filtering in a linear MT chain.** For the linear MT chain, the MT triangle and for comparisons with single filaments, we analyze the full oscillation period of the anchor points leading to compression and stretching of the filaments. The serial connection of two 10 µm microtubules (stabilized by 100 µM Taxol) held by three optically trapped anchor points (bead $i = 1$, beads $j = 2, 3$) revealed an unexpected elastic behavior. Relative to the first bead connection 1→2 with one MT, the addition of a second MT makes the new bead connection 1→3 five times stiffer in longitudinal direction and five times softer in lateral direction. Hence, longitudinal momentum can be well transported through this linear construct, but lateral momentum is damped such that the linear construct acts as angular filter for the transport of mechanical momentum. Remarkably, the single filament description of the elastic modulus $G'(\omega)$ as the sum of a frequency independent part $G'(0)$ and a part following a power law is still valid and the transition frequency for longitudinal momentum transport also increases with increasing $G'(0)$. Since longitudinal and lateral tubulin bonds differ in strength, one can conclude that the geometry of the network and the angular direction of momentum transport affect the molecular relaxation behavior of individual tubulin heterodimers and the stabilizing molecules bound to these heterodimers.

The two-step elastic modulus $G'^{(1,3)}$ can be modelled as serial connection of two springs, resulting in an additional coupling term $G_{cpl}$ as explained in the Methods section. It is likely that $G'_{cpl}(\omega, q, F_{ext})$ depends on an external force $F_{ext}$, which is given in our case by the optically trapped bead 2. The stronger this external (optical) force, the less lateral oscillations can be transferred from the first to the second microtubule, and the stiffer is the connection in longitudinal direction. This trapping effect has been quantified for different stiffnesses of the optical trap as shown in the Fig. 4d, but has not yet been subtracted to obtain the pure elasticity of the filament itself. In general, we have no control of the attachment of both filaments to the second, intermediate bead 2. They might attach perfectly opposite to each other or nearly at the same location giving rise to different effective suspensions of the middle bead.

The role of the intermediate (trapped) bead simulates the situation inside a cell, where filaments are cross-linked to each other, and to other cellular components, such as actin. These crosslinkers have different elasticities, hence, we can test the situation *in vivo* by varying the trap stiffness of the intermediate bead. A recent theoretical study investigated the role of crosslinkers in reversibly crosslinked networks of semi-flexible polymer filaments and found a qualitatively similar behavior, i.e., a low frequency plateau depending on the number of crosslinkers and a power law rise at high frequencies [49].

Interestingly, this situation resembles an (electronic) transistor, where a small input signal (here, a mechanical stimulus) controls a strong current (here, the momentum transport from bead 1→3). It will be interesting to perform further experimental and theoretical investigations to explain the elastic behavior, where momentum transport between two network nodes can be steered by an intermediate node.

**The MT triangle – a uni-directional stable network.** Displacement of the actor bead in either radial x- or tangential y- direction as illustrated in Fig. 5 results in a very direct and efficient transport of momentum in direction towards the one or the other sensor bead. Remarkably, the measured elasticity behavior described by the modulus $G'(\omega)$ is the same as in the single filament case. It consists of a static elasticity $G'(0)$, and a strong rise of $G'(\omega)$ when higher deformation modes are excited beyond the transition frequency $\omega_t$. This strong rise is clearly visible at $f_t = 100$ Hz for a tangential oscillation, but could not be resolved for a radial oscillation. This is probably due to a much faster rise (larger exponent) of $G'$ in a direction lateral to the filament axis, as we observed this phenomenon for single filaments and the linear MT chain as well. However, based on our observations for single filaments, we could estimate the transition frequency for a radial oscillation of the triangle to be $f_t \approx 800$ Hz. In the static case, the triangle is about 25 times stiffer than a single filament. This can be explained by the fact that every radial or tangential displacement of the actor bead results in a compression and stretching of another MT at the same time. Since MTs are hardly stretchable [32], this results in static elasticities of $G'(0) \approx$ 20-50 mPa. Comparing the estimates for the transition frequency $\omega_t$ obtained from $G'(0)$ and equation (3), these extrapolated values come close to the frequency where $G'(\omega) \approx 1.5 \cdot G'(0)$. Again, we interpret the increased transition frequency as a result of the intermolecular relaxations of or between two tubulin heterodimers, which cannot follow on timescales below $2\pi/\omega_t < 10$ ms. A stiffening beyond a transition frequency of $f_t \approx 200$ Hz could also be observed in cross-linked actin networks [50]. Whereas the optically trapped anchor points could rotate and act as hinges in the previous configurations, the anchor points of the triangle can hardly rotate, and therefore rather resemble a movable support only. This triangular situation is relevant for the radial MT arrays that form around the nuclei of many cells by a mechanism where microtubule-nucleation factors are directionally transported by dynein motors [51]. In addition, the forces conveyed to the nucleus by this network would act, via links of the cytoskeleton to the nuclear lamina, on structure and dynamics of the chromatin [52], providing a mechanism how mechanic signals can modulate gene activity in the network's center.

**Summary and conclusions**

Motivated by the capability of individual microtubules and inter-connected microtubule networks to transduce a mechanical stimulus over a long distance within short times, we clearly identified substantial differences in response for different network topologies and at different stimulation frequencies $\omega$ [35]. This has a couple of interesting implications for biology:

The rather low stiffness at frequencies below the characteristic transition frequency, $\omega < \omega_t$, of single filaments or the linear network is expected to dampen the transmission of mechanical signals, while the rise at $\omega > \omega_t$ would allow for an enhanced transmission of signals that typically show a reduced amplitude in noise driven systems such as living cells. Interestingly, this transition frequency is in a physiologically relevant range (1-10 Hz). For instance, the mammalian heartbeat ranges between 1 Hz in humans up to 18 Hz in mice [53], and muscles undergo an innate oscillation of around 20 Hz [54].

A second aspect of the strong influence of network topology is the comparatively high stiffness at low frequencies of triangular networks. This displays a stiff, load bearing scaffold, which could be used to reinforce the cell against external pressure in densely packed tissues, or enable the contraction of large scale MT networks [55]. The specific mechanical properties of triangular networks are relevant for nuclear positioning, since the nucleus is tethered and positioned by radial arrays that are stabilized by cross-connection in many organisms integrated into cell polarity [56]. A third implication of our findings is linked with the "mechanic transistor" function of microtubule networks, where small mechanical forces can control a large amount of momentum transport.

Microtubule crosslinkers have recently been reported to be able to generate entropic forces on the pN range [57], which could lead to passive changes of network elasticity over time by pre-stretching individual filaments of a network. This would provide a mechanism how cells can control the directionality of mechanic signaling, which is relevant for mechanic integration of cells into organs, or of organs into organisms [2]. These implications show that our bottom-up

Page 15

approach to analyze the transmission of mechanic forces in networks of increasing complexity is relevant to understand, how mechanic signals can shape biology.

**Materials and methods**

**Theoretical description of viscoelastic behavior**

This section introduces the relevant forces acting on a single filament and its resulting deformations as well as the relative bead displacements during an oscillation longitudinal to the MT. Through a representation of all forces in frequency space, the elastic and viscous components of the filament can be extracted using the frequency dependent complex shear modulus $G(\omega)$.

To separate the viscoelastic contributions of the filament and the trapped beads, we analyzed the data by means of active two particle micro-rheology in frequency space. Because the microtubule is firmly attached to the beads, every displacement $x_B$ of a bead in x direction directly results in an evasion of the microtubule, i.e., buckling with amplitude $u(x, x_B)$. Hence, the forces acting on the microtubule and the forces on the beads are directly coupled through the constraint of a constant contour length $L$. As we show in the Supplementary Results, the measured net forces on the beads in direction lateral (y) to the filament are negligibly small, such that all effective forces due to microtubule buckling and viscous drags point only in x direction. Hence, in the tension free case the sum of forces acting on a single bead with index $j$ can be described by the following, one dimensional equation of motion for the bead at longitudinal position $x_{Bj}$ and the filament contour described by $u(x)$:

$$F_{opt\,j}(x_{Bj}) + F_{\gamma_{Bj}}(x_{Bj}) + \tfrac{1}{2} F_{\kappa_{MT}}(u(x)) + \tfrac{1}{2} F_{\gamma_{MT}}(u(x)) = F_D \quad (4)$$

Here, $F_{opt\,j} \approx -\kappa_T \left( x_{Bj}(t) - x_{Lj}(t) \right)$ is the elastic optical force and $F_{\gamma_{Bj}} \approx -\gamma_B \frac{\partial}{\partial t} x_{Bj}(t)$ with $\gamma_B = 6\pi R_B \eta$ the translational viscous drag force both acting on bead $j$ and balancing the counteracting elastic buckling force $F_{\kappa_{MT}} = \int_0^L EI \cdot \left| \frac{\partial^4}{\partial x^4} u(x,t) \right| dx$ [33] and the viscous drag force $F_{\gamma_{MT}} \approx \int_0^L g_{MT} \cdot \frac{\partial}{\partial t} |u(x,t)| dx$ of the MT filament integrated along the contour length $L$ and with bending modulus (flexural rigidity) $EI = k_B T \cdot l_p$, where $l_p$ is the persistence length of the semiflexible polymer with $L << l_p$. $g_{MT} = \frac{4\pi\eta}{\ln(L/D)+0.84}$ is the lateral viscous drag coefficient per unit length, and $\eta$ the viscosity of water [58]. Both in our theoretical description as well as in our rheological analysis shown in Figs. 3, 4, and 6, we only include the effects of buckling during single filament compression, and neglect microtubule stretching because the microtubule stretching spring constant is on the order of 10 pN / nm [32]. This would result in a maximal extension during our experiments of approximately 1 nm, compared to large buckling deformations on the order of several 100 nm. The absolute values in the expressions for the total buckling and viscous forces of the filament are due to symmetry: filament buckling in positive or negative direction $\pm u(x)$ must always result in the same force on the beads.

The oscillatory driving force $F_D(t) = -\kappa_T \cdot x_{L1}(t)$ is generated by the first optical trap ($j=1$, actor trap) at position $x_{L1}(t) = A_1 \sin(\omega_a t)$, thereby compressing and stretching the MT. The equation of motion according to equation (4) can then be given explicitly:

$$-\left(\kappa_T + \gamma_B \tfrac{\partial}{\partial t}\right) x_{Bj}(t) \pm \tfrac{1}{2} \int_0^L \left( EI \left|\tfrac{\partial^4}{\partial x^4} u(x,t)\right| + g_{MT} \tfrac{\partial}{\partial t} |u(x,t)| \right) dx = -\kappa_T A_j \sin(\omega_a t) \qquad (5)$$

Hence, equation (5) represents a set of $m$ coupled differential equations, where $m$ is the number of beads. These are solved pair wise using relative and collective coordinates $x_R = x_{B1} - x_{B2}$ and $x_C = x_{B1} + x_{B2}$. Since the contribution of the filament acts in opposite directions for each bead (points away from the MT ends, $\pm$ in equation (5)), this effect cancels out in the collective coordinate $x_C(x_{B1}, x_{B2})$, but manifests in the relative coordinate $x_R(x_{B1}, x_{B2}, u)$. The buckling of the filament contour $u(x_{B1}, x_{B2}, x, t)$ is a function of the compression given by the bead positions $x_{B1}$ and $x_{B2}$ and is assumed to be deformed in lateral direction $y$ only with small angles to the x-axis. The deformation amplitude can be written as a superposition of sinusoidal modes with wavenumber $q_n = n \cdot \pi / L$ ($n \geq 1$) [59, 60, 61]:

$$u(x,t) = \sum_{n=1}^{N} u_{qn}(t) \cdot \sin(q_n x), \qquad (6)$$

The amplitudes $u_{qn}(t) = \tfrac{1}{n\pi}\sqrt{2L\delta_L(t) - \delta_L^2(t)}$ of these modes decay exponentially with time according to the temporal auto-correlation function $AC[u(x,t)] = \tfrac{L}{2q_n^4 \ell_p} \exp(-\omega_{qn} t)$ and can be estimated considering the constant arc length $L = \int_0^{L-\delta_L} \sqrt{1 + \left(\tfrac{\partial}{\partial x} u(x,t)\right)^2} \, dx$ of the buckled MT. The filament is axially compressed by $\delta_L(t) \approx x_{L1}(t)$, which is a reasonable approximation to the resulting elliptic integral, as shown in the Supplementary Results. The mode relaxation with frequency $\omega_{q_n} = \tfrac{EI}{g_{MT}} q_n^4$ is the faster, the larger the wave number and the bending modulus $EI$. For a single deformation mode $q_n$, equation (5) then becomes for the relative coordinate $x_R$

$$-\left(\kappa_T + \gamma_B \tfrac{\partial}{\partial t}\right) \cdot x_R(t) + \left(EI \cdot q_n^4 + g_{MT} \tfrac{\partial}{\partial t}\right) \cdot u_{qn}(t) \int_0^L |\sin(q_n x)| dx = F_D(t) \qquad (7)$$

with $\int_0^L |\sin(q_n x)| dx = n \int_0^{L/n} \sin\left(\tfrac{n\pi}{L} x\right) dx = \tfrac{2 \cdot n}{q_n}$. Using the temporal frequency $\omega$ and the Fourier relation $\tfrac{\partial}{\partial t} \to i\omega$, equation (7) reads in frequency space:

$$-\left(\kappa_T + i\omega\gamma_B\right)\cdot\tilde{x}_R(\omega) + \left(2EI\cdot nq_n^3 + 2i\omega g_{MT}\tfrac{n}{q_n}\right)\cdot\tilde{u}_{qn}(\omega) = \tilde{F}_D(\omega) \qquad (8)$$

The spectral forces acting on the beads with relative position $\tilde{x}_R(\omega)$ are known and can be subtracted, such that the following response equation holds: $\tilde{u}_{qn}(\omega) = \alpha_{qn}(\omega)\Delta\tilde{F}_D(\omega)$. Here, $\Delta\tilde{F}_D(\omega)$ is the Fourier transform of the external force $F_D(t) + \left(\kappa_T + \gamma_B \tfrac{\partial}{\partial t}\right)x_R(t)$, which deforms the MT at different temporal and spatial frequencies.
$\alpha_{qn}(\omega) = \left(2EI\cdot nq_n^3 + 2i\omega g_{MT}\tfrac{n}{q_n}\right)^{-1}$ is the end-to-end response function of a single microtubule deflected by $\tilde{u}_{qn}(\omega)$ upon $\tilde{F}_D(\omega)$, which simplifies to $\alpha_{qn}(\omega) = \left(2EI\right)^{-1}\cdot\left(nq_n^3 + inq_n^3\tfrac{\omega}{\omega_n}\right)^{-1}$
$= \left(2kT\cdot\ell_p nq_n^3\right)^{-1}\cdot\left(1 + i\tfrac{\omega}{\omega_n}\right)^{-1}$, with units $[\alpha_{qn}] = \tfrac{m}{N}$. The total response function over all $N$ deformation modes can be calculated as a superposition of $N$ individual response functions $\alpha(\omega) = \sum_n \alpha_{qn}(\omega)$. Using the wave number $q_1 = \tfrac{\pi}{L} = \tfrac{1}{n}q_n$ and the relaxation frequency $\omega_1 = \tfrac{EI}{g_{MT}}q_1^4 = \tfrac{1}{n^4}\omega_n$ of the ground mode with $n = 1$, one obtains:

$$\alpha(\omega) = \tfrac{1}{2q_1^3 kT\ell_p}\sum_{n=1}^{N}\tfrac{1}{n^4 + i\omega/\omega_1} \qquad (9)$$

with $\alpha(0) = \tfrac{1}{2q_1^3 kT\ell_p}\sum_{n=1}^{\infty}\tfrac{1}{n^4} = \tfrac{\pi^4}{180 q_1^3 kT\ell_p} = \tfrac{0.54}{q_1^3 kT\ell_p}$. The complex viscoelastic response functions of the MT filament consists of the storage modulus $G'(\omega)$, describing the elastic energy stored in the system, and the loss modulus $G''(\omega)$, describing the friction energy dissipated to the environment. By using $G(\omega) = \tfrac{1}{4\pi L\cdot\alpha(\omega)}$, we find :

$$G'(\omega > \omega_1) = \text{Re}\left\{\tfrac{1}{4\pi L\cdot\alpha(\omega>\omega_1)}\right\} \approx C\cdot\omega^p \qquad (10)$$

$$\tfrac{1}{1.86}G'(\omega_1) = G'(\omega = 0) = \tfrac{1}{2.16\pi^2}q_1^4 k_B T\ell_p = \tfrac{1}{2.16\pi^2}g_{MT}\omega_1 \qquad (11)$$

and $G''(\omega) = \tfrac{\text{Im}(\alpha(\omega))}{|\alpha(\omega)|} \approx \tfrac{\eta\omega}{\ln(L/D)+0.84}$ where $C$ is a constant factor. For higher driving frequencies $\omega > \omega_1$, $G'(\omega)$ follows a power law with $p = 3/4$ [59], whereas a more advanced theory for semiflexible filaments [36, 37] predicts a power law with $p = 5/4$. For low frequencies $\omega \to 0$, the elastic modulus is close to the first mode $G'(\omega_{n=1})$, which is independent of the frequency. In the following, we only investigate the elastic component $G'$, whereas the viscous contributions $G''$, expressed by the viscous drag $g_{MT}$ of the MT, are discussed in the Supplementary Results.

**Theoretical estimate for MT stiffening on short timescales**

The question is how well our observations can be explained on the basis of an equation of forces, as introduced in equation (4), and viscoelastic forces known from hydrodynamic beam theory. Our theoretical description of microtubule deformation through the shear modulus $G'(\omega)$ is based on the beam equation $M(x) = EI \frac{d\theta(x)}{dx}$ with bending moment $M$ and tangent angle $\theta(x)$ along the filament [58], which has been successfully applied to active filament stretching [32] and buckling [31] and to thermal deformations [33]. On this basis, the static MT deformations result in a strong length dependency of $G'(\omega=0) \sim \left(\frac{1}{L}\right)^4$. However, by considering the contour-length dependence of the persistence length $l_p(\omega = 0, L) \sim L^2$ [38], the plateau value $G'(0)$ as well as the transition frequency $\omega_t$ approximate to a $\left(\frac{1}{L}\right)^2$ dependency. Whereas this could be confirmed for the plateau value, the description of the transition frequency requires a more advanced theory, which should also include the molecular architecture of differently stabilized filaments.

As introduced above, the buckling amplitude of the filament deformation $u(x_{B1}, x_{B2}, t)$ is a superposition of different deformation modes, which relax the faster, the larger the wave number $q_n = n \cdot \pi/L$, or the shorter the deformation length. By calculating individual response functions $\alpha(\omega, q)$ for each mode $q_n$ and taking the inverse sum of all response functions, $G'(\omega, n) = 1 / (\Sigma_n \alpha(\omega, q))$, both the elastic and viscous modulus are obtained. It turned out that typically $n = 4$ modes were excited at our maximum driving frequency of 100 Hz, such that the fit function $G'^{(4)}_{fit}(\omega) = \frac{3(59\omega^6 + 1383427\omega_1^2\omega^4 + 1520884952\omega_1^4\omega^2 + 19790659584\omega_1^6)}{2A(4\omega^6 + 176037\omega_1^2\omega^4 + 511196337\omega_1^4\omega^2 + 32023818304\omega_1^6)} \approx G'(\omega, n=4)$ with the parameters $A$ and $\omega_1$ was sufficient.

Using this fit function, the transition frequency $\omega_t \approx 3\omega_1$ could be extracted from the experimental data. $\omega_t$ was interpreted as the frequency at which molecular relaxations cannot follow the external filament deformation. The frequency independent stiffness at low frequencies and the sudden increase in $G'(\omega)$ on a double-logarithmic scale could be well observed in single filaments as well as in the linear and triangular MT arrangements. From these observations, we conclude that the description of forces chosen in equations (4) and (5) to quantify our mechanistic model is reasonable. However, the stronger stiffening at high frequencies with p > 5/4 needs a more thorough theoretical investigation. In addition, the theoretical approach has to be extended in the future, to also integrate the porous molecular structure, especially to explain the dependence of the transition frequency on chemical stabilization of the microtubule (see Supplementary Results).

**Stiffness estimate for a linear MT chain**

The two-step elastic modulus $G'^{(1,3)}$, resulting in a fivefold stiffening in longitudinal direction and fivefold softening in lateral direction compared to the one-step modulus $G'^{(1,2)}$, can be modelled as serial connection of two springs (two filaments, $2fl$) with MT length $L/2$ or wave number $2q$, such that $G'^{(1,3)} \to G'_{2fl}(2q)$. Reciprocal addition of two single filament elasticities $G'_{1fl}(2q)$ results in a two filament sum elasticity, $G'_{2fl}(0,2q) = \left(\frac{1}{G'_{1fl}(0,2q)} + \frac{1}{G'_{1fl}(0,2q)}\right)^{-1} = \frac{1}{2}G'_{1fl}(0,2q) \neq G'^{(1,3)}(0,2q)$, which is two times softer than that of a single filament. Alternatively, the two-step modulus $G'^{(1,3)}$ can be identified with a single filament of length $L$, or wave number $q$, such that $G'^{(1,3)} \to G'_{1f}(q)$. However, this results in a fivefold decrease of the elasticity relative to that of a single filament with length $L/2$, according to $G'_{1f}(0,2q) = l_p(0,2q)k_BT(2q)^4 = \frac{16}{3}l_p(0,q)k_BTq^4 \approx 5 \cdot G'_{1f}(0,q)$. The factor 1/3 arises from the length dependence of $l_p(q)$.

Hence, an additional coupling term $G'_{cpl}$ is required to explain the elastic behavior of the linear construct, such that

$$G'^{(1,3)}(\omega, q) = G'_{1f}(\omega, q) + G'_{cpl}(\omega, q) \tag{12}$$

$G'_{cpl}(0,q)$ must be positive for longitudinal momentum transport, and negative for lateral momentum transport.

**Experimental setup with optically trapped beads as actor and sensor**

A single biotinylated microtubule was attached laterally to two Neutravidin coated beads trapped by time-multiplexed optical tweezers and aligned along the x-direction as illustrated in Fig. 1. The average laser power per trap was always 19 mW in the focal plane at a trapping wavelength of 1064 nm. The position of each bead was tracked in three dimensions at 50 kHz using back focal plane (BFP) interferometry and quadrant photo diodes (QPD) as described in [62]. The first optical trap (trap 1, shown in red) was the force generating actor and oscillates sinusoidally at frequency $f_a$ and amplitude $A_a$ along $x$ around the central position $x_{01}$. The other trap(s) (blue) remained static and were used as position and force sensors for the mechanical stimulus exerted by the actor and transduced by the microtubule. The beads were displaced by $x_{B1}$ and $x_{B2}$ relative to the trap centers. During both half periods of an oscillation, the distance between the beads was first increased and then decreased resulting in tensile and compressive forces acting on the microtubule, respectively. Since microtubules are practically inextensible, they are bent locally at the point of attachment to beads (Fig. 1a) during the first half period [32] and buckled during the second half period due to their high compliance to compression forces [63]. The buckling amplitude along the filament is denoted by $u(x, t)$ as illustrated in Fig. 1c.

In the experiments, we used a lateral stiffness of $\kappa_{opt} \approx 25$ pN/μm per trap. The actor trap was typically oscillating at frequencies 0.1 Hz $\leq f_a \leq$ 100 Hz in nearly logarithmical steps and amplitudes 200 nm $\leq A_a \leq$ 600 nm along $x$ or $y$, i.e., longitudinal or lateral with respect to the axis of single filaments or the linear chain. Each filament or construct was probed several times at different amplitudes to test for any force or displacement dependence, to obtain statistics, and to test for structural defects during experiments, which happened rarely and usually resulted in filament breaking close to one bead. In such cases, the measurements were excluded from further analysis. Also, we did not observe significant differences for repetitive measurements on the same filament indicating that microtubules were not structurally damaged during oscillation. In some cases, one of the filaments was detached of a bead. These experiments have also been excluded from analysis. Elastic effects of the biotin linker can be neglected, since the effective length of this linker is in the Ångstrom range [64] and its spring constant [65, 66] is much larger than that of microtubules, both for buckling and stretching [32].

**Suitability of experimental approach**

The use of optically trapped beads as anchor points for simple microtubule networks turned out to be a very suitable approach. Potential phototoxic effects such as bleaching and filament breaking were successfully suppressed by addition of glucose oxidase and catalase as enzymatic scavengers of reactive oxygen species (see Microtubule preparation), which allowed to obtain reproducible results for more than 30 experimental repeats on the same construct extending up to 40 minutes. At a moderate laser power of 19 mW per trap, optical forces were high enough to compensate all occurring friction and elastic forces. Only for trap displacements at frequencies strongly exceeding 100 Hz, the trapping of higher refracting polystyrene spheres became unstable [67]. However, most physiologically relevant mechanical forces on timescales below (100 Hz)$^{-1}$ = 10 ms are local effects on the order of one pN or less, caused by thermal fluctuations and are likely not relevant for a transport through the entire cell. As we have shown in the Supplementary Results, the extents of the 1µm large beads result in additional geometrical effects affecting the deformation and have to be considered in the future. However, these effects are minor and do not impair the feasibility of the strategy to use individually trapped beads as flexibly controllable force actors and sensors within small MT networks. The basis for all experiments were the multi-particle trapping, the precise position tracking and the force measurement in well calibrated optical traps, which worked robustly at the used tracking rates of 50 kHz / $N$ ($N$ = number of traps). We think that this successful approach for optical trapping and tracking of anchor points will also allow investigating more complicated networks in the future.

**Microtubule preparation**

Tubulin was purified from fresh brains collected freshly after slaughtering using the classical protocol by Shelanski et al. [68]. For biotinylation, microtubules were preassembled at 37°C in presence of 100 µM taxol and 100 µM of GTP in BRB80 buffer (80 mM Pipes KOH, pH 6.8, 1 mM $MgCl_2$, 1 mM EGTA), and then complemented with 500 µM of sodium bicarbonate and 1 mg/ml of biotin-XX N-hydroxysuccinimide ester. After incubation for 30 min at 37°C, the mixture was purified by ultracentrifugation through a twofold volume of a sucrose cushion (15 min 300000 g) in BRB80. Purity and quality of each tubulin preparation was verified by SDS-PAGE, before coupling the purified tubulin to tetramethyl rhodamine as described previously [69]. For Taxol stabilized microtubules, tubulin, fluorescently labeled tubulin, and biotinylated tubulin were thawed on ice, mixed with GTP in the ratio 8 : 4 : 4 : 0.8 and polymerized for 30 min at 37°C. This stock was stable up to 2 days at room temperature. Dilutions (1:100 – 1:2000) in BRB80 buffer containing Taxol (10 µM or 100 µM) were prepared freshly from the stock every 2-3 hours during experiments. Doubly stabilized filaments were polymerized similarly, but with Guanosine-5'-[(α,β)-methyleno]triphosphate (GMPCPP) instead of GTP and spun down with a TLA100 rotor in a Beckman centrifuge at 300000g. Sedimented microtubules were resuspended in BRB80 buffer containing 100 µM Taxol and stable for 2-3 months at room temperature. Further dilutions (1:100 – 1:500) were prepared freshly during experiments. 7.5 µl of the microtubule suspension was mixed on a coverslip with Neutravidin coated beads (Molecular probes, Invitrogen, F8777), and an oxygen scavenging system (GODCAT, 100µg/ml Glucose oxidase 22778 from Serva, 20µg/ml catalase C40 from Sigma, 10mM BME M3148 from Sigma and 40mM Glucose from Carl Roth) to prevent fluorescence bleaching and filament breaking. We always used a roughly 80μm thick coverslip sandwich separated by double sided sticky tape (Tesa).

**Acknowledgements**

The authors gratefully acknowledge helpful discussions with Igor Kulic, Falko Ziebert, and Felix Jünger as well as Stefan Diez and Friedrich Schwarz for providing GMPCPP filaments. The project was funded by the Deutsche Forschungsgemeinschaft (DFG), grant RO 3615/2-1 and RO 3615/2-3.

**Author contribution**

M.K. designed and built the optical setup, performed experiments, analyzed data and prepared figures. N.S. and P.N. contributed microtubules and reagents. P.N. commented on the manuscript and contributed to the Discussion and interpretations of the results. A.R. developed the theory with M.K. M.K. and A.R. wrote the paper. A.R. initiated and supervised the project and obtained financial funding.

**Competing Financial Interests statement**

The authors declare no competing financial interest.

**Figure legends**

Fig. 1. Design of experiment illustrating different stages of an oscillation period. Optically trapped $d = 1.06$ µm Neutravidin-coated beads are used as anchors points for laterally attached, biotinylated microtubules. Bead positions are tracked interferometrically by QPDs. (a) The filament is stretched during the first half oscillation period. (b) The filament is straight after each half period. (c) The filament is buckled during the second half oscillation period. Insets: overlay of corresponding fluorescence and brightfield images.

Fig. 2. Frequency dependent response of a single microtubule filament – bead construct (L = 5µm). (a) and (b) Two periods of the relative actor and sensor bead displacements $x_{B1}(t) - x_{L1}(t)$ and $x_{B2}(t) - x_{L2}(t)$ from their trap centers during oscillations of the actor trap $x_{L1}(t) = L/2 + A_a\sin(\omega_a t)$ and the static sensor trap $x_{L2}(t) = -L/2$ at $A_a = 500$ nm and $\omega_a/2\pi = 0.1$ Hz (a), or $\omega_a/2\pi = 100$ Hz (b), respectively. Markers indicate the maximum amplitude $x_{max1}$ and $x_{max2}$ of both beads during each half period. (c) Frequency dependence of the maximum amplitude $\Delta L_x = A_a - x_{max1} - x_{max2}$ between both beads during stretching and buckling averaged over several oscillation periods. (d) Theoretical mean amplitudes of the involved viscous forces (solid lines) compared to the experimentally obtained sum of all acting forces (markers) for a single trapped bead (yellow) and the filament – bead construct (red).

Fig. 3. Elastic modulus for differently stabilized single microtubules of different lengths. (a) Stabilization with 10 µM Taxol, MT end oscillation direction: longitudinal. N = 8 measurements for short and N = 19 measurements for long microtubules. (b) Stabilization with 100 µM Taxol, MT end oscillation direction: longitudinal. N = 9 measurements each length. (c) Stabilization with GMPCPP + 100 µM Taxol, MT end oscillation directions: longitudinal (solid black line) and lateral (dashed black line). N = 9 measurements each direction. (d) Frequency dependent persistence lengths obtained from the elastic moduli for all filament types. Error bars represent the standard deviation (SD).

Fig. 4. Rheology of a linear chain of connected filaments with transistor function. (a) Experimental design of linear network of two filaments held by three trapped beads. Bottom: overlay of fluorescence and brightfield image. Scale bar: 5 µm. (b) and (c) Elastic components G' for a longitudinal and lateral MT end oscillation, i.e., in directions longitudinal or lateral to filaments, respectively. The transition frequencies are marked by black dotted lines. (d) Elastic components $G'(1,3)$ for a longitudinal oscillation and different stiffnesses $\kappa_2$ of the intermediate optical trap. Error bars represent the standard deviation (SD) of N = 3 measurements each.

Fig. 5. Probing an equilateral triangular network of GMPCPP filaments. (a) Design of experiment in pseudo 3D with overlay of fluorescence and brightfield image (bottom). Scatterplots of absolute bead positions during a tangential oscillation in $y$ direction are shown color coded over the current phase $\phi_a$ of the actor trap. (b) Overlay of fluorescence and brightfield images of the characteristic time points during one radial (x) oscillation period at $f_a = 0.1$Hz. Scale bar: 5µm.

Fig. 6. Elastic components $G'(\omega)$ for an equilateral triangular network. These are compared to the theoretical estimates for single filaments (lines without markers). (a) Radial oscillation along *x*. (b) Tangential oscillation along *y*. Error bars represent the standard deviation (SD) of N = 10 measurements each.

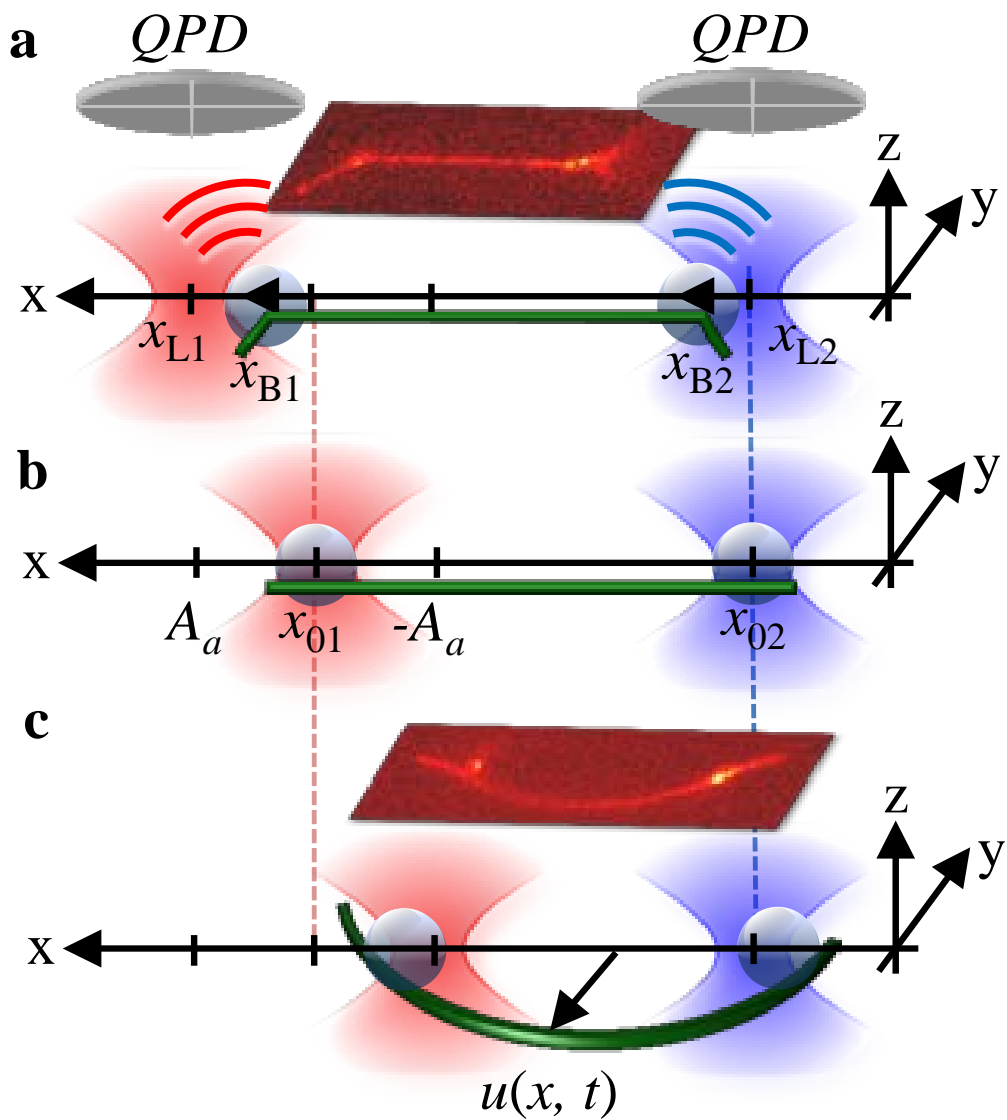

Trap 1: oscillating
Trap 2: static

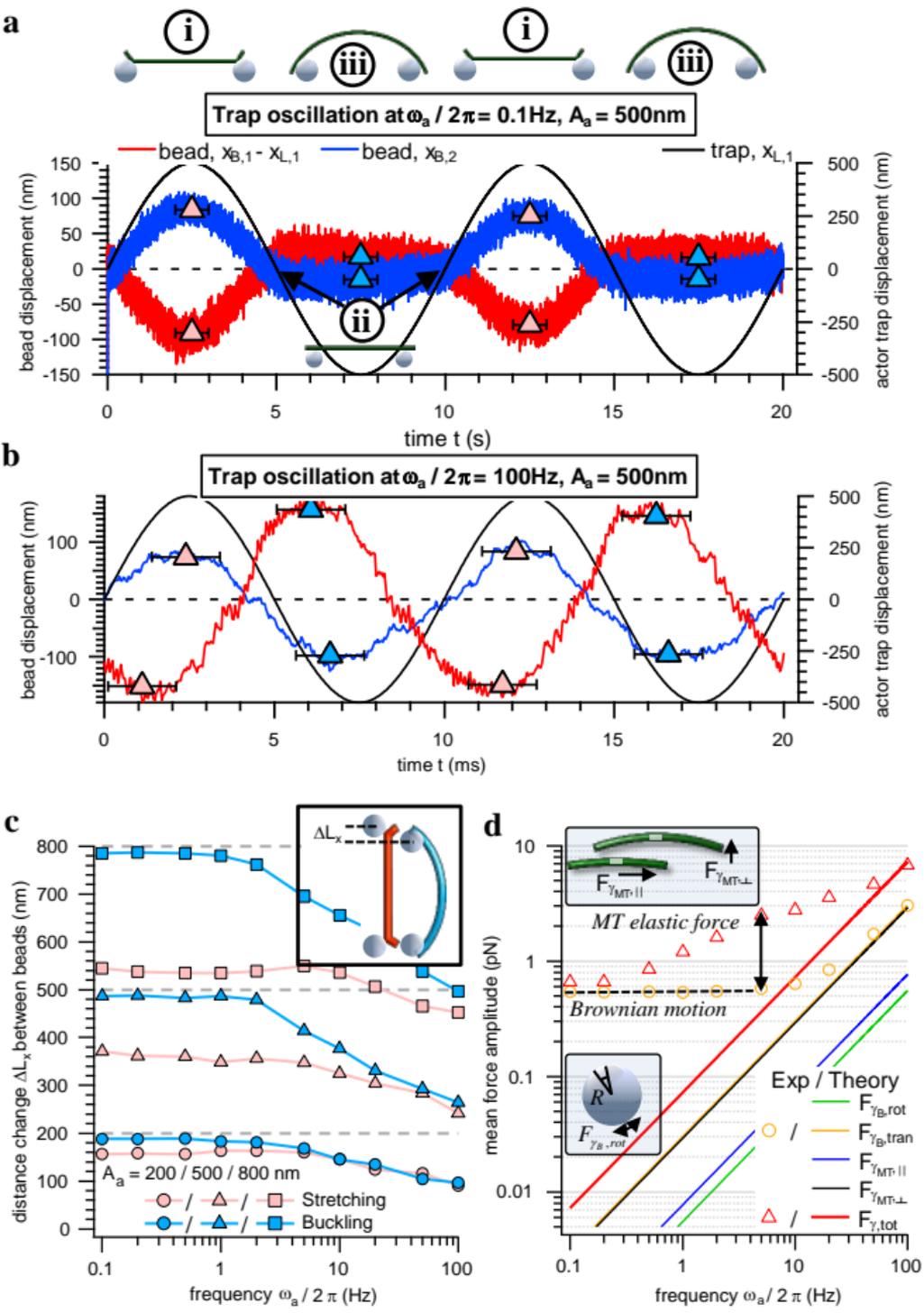

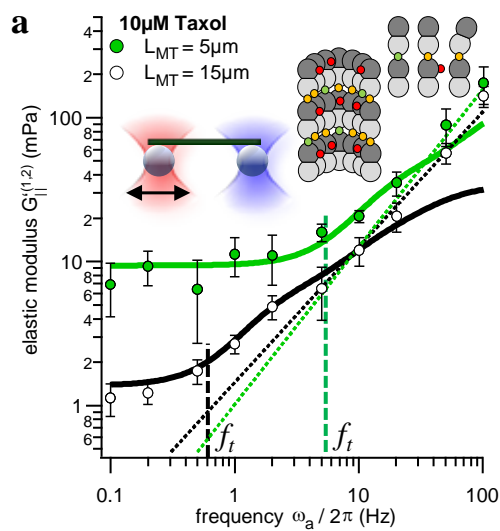
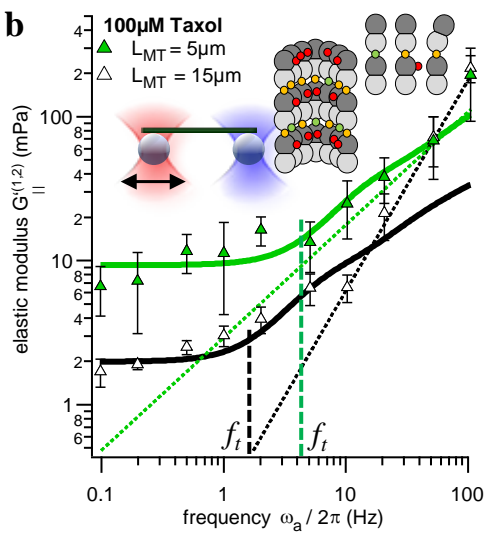
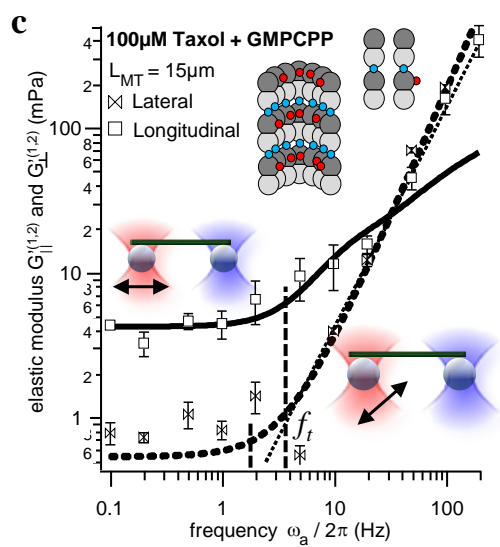
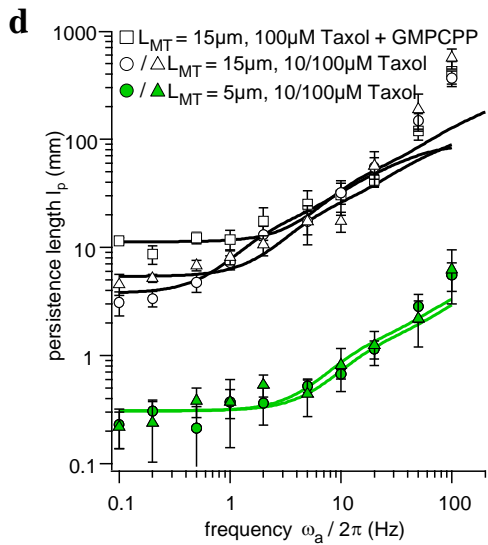

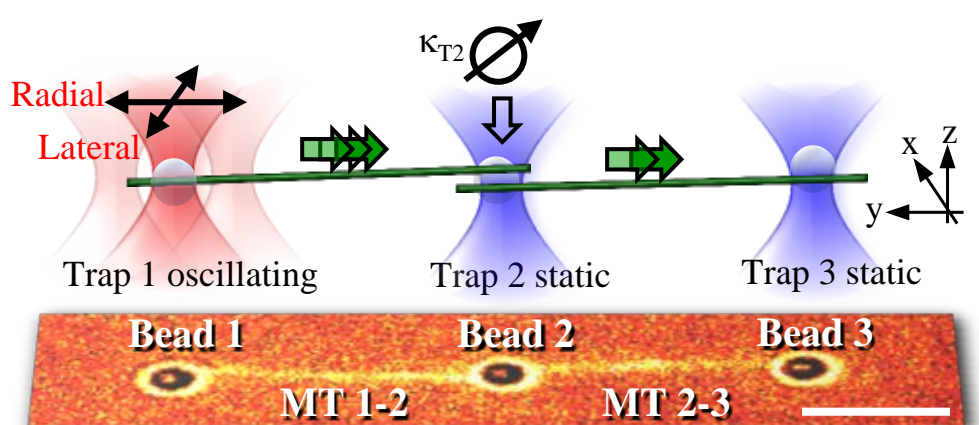
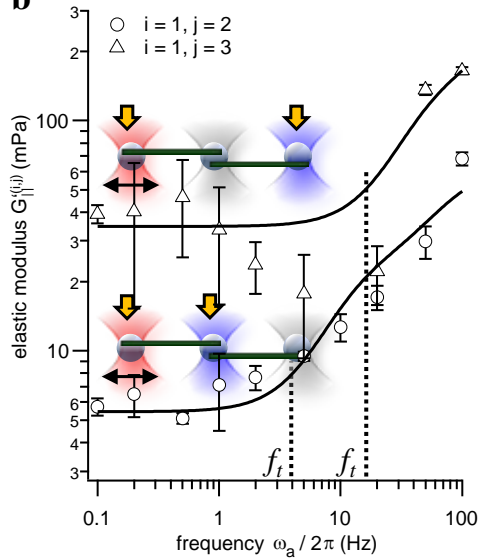
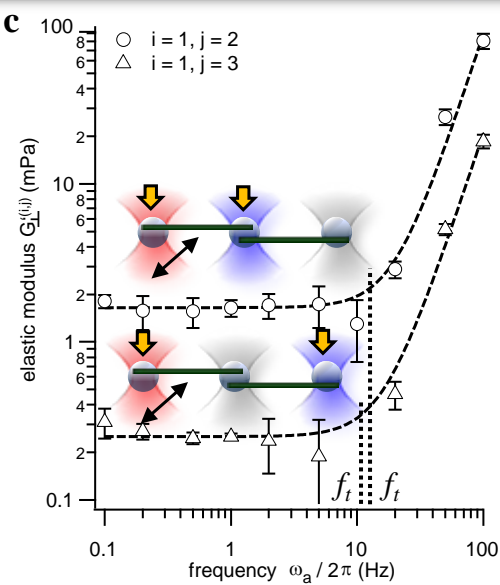
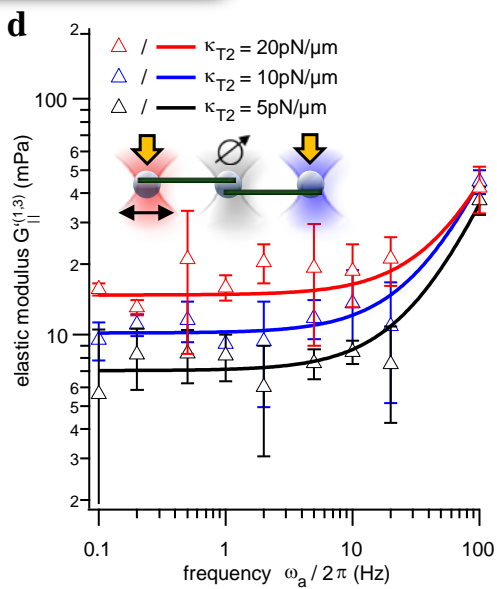

**a**

Trap 1 oscillating
Trap 2 static
Trap 3 static

15 μm
15 μm
15 μm

Bead 1
Bead 2
Bead 3

MT 1-3
MT 1-2
MT 2-3

Color bar

$\phi_a$ ($2\pi$)

**b**

$t = 0$ — Relaxed, Relaxed, Relaxed

$t = T/4$ — Tense, Tense, Buckled

$t = T/2$ — Relaxed, Relaxed, Relaxed

$t = 3/4T$ — Buckled, Buckled, Tense

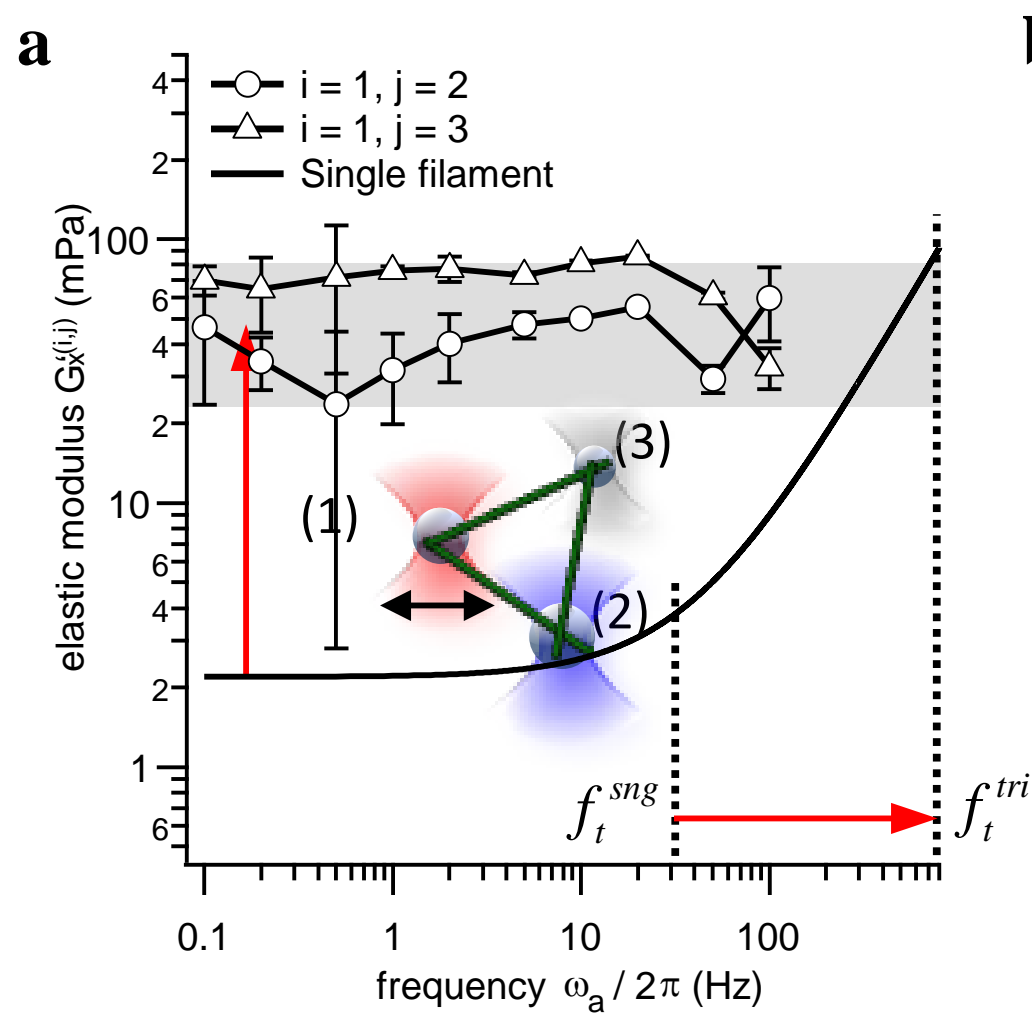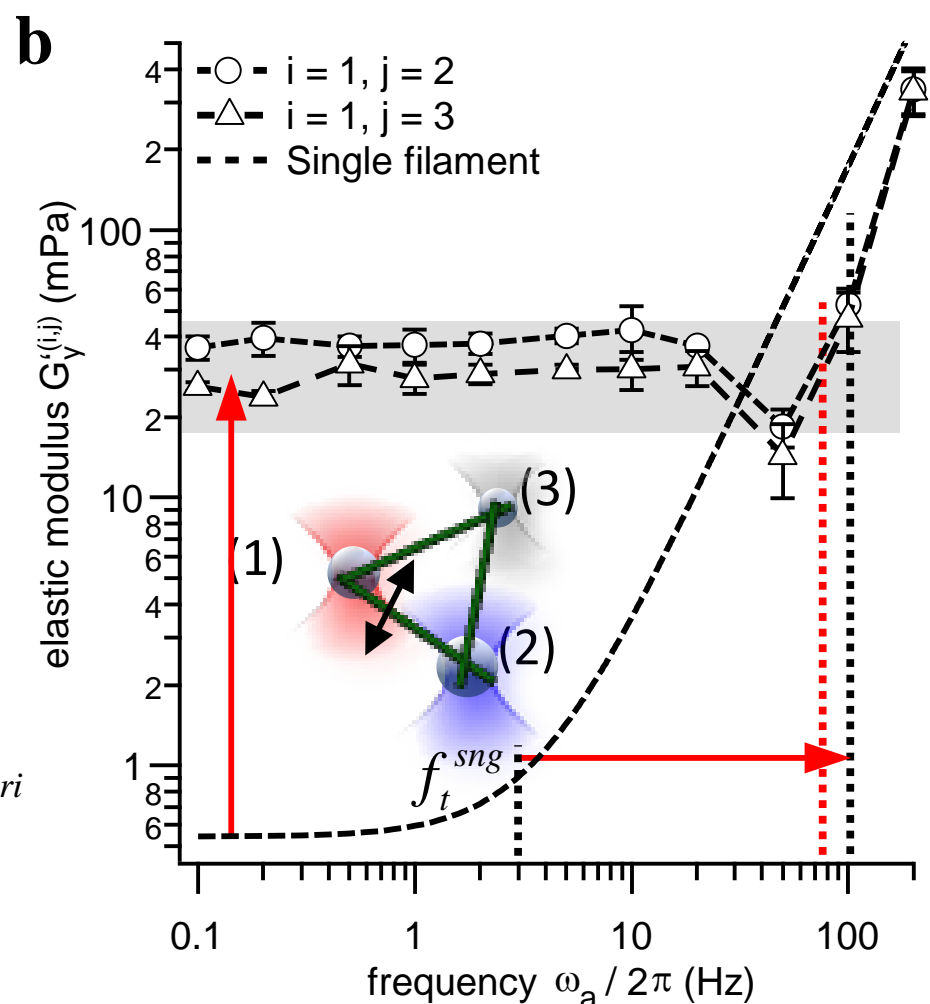

# Single microtubules and small networks become significantly stiffer on short time-scales upon mechanical stimulation

## Supplementary information

Matthias D. Koch[a,1], Natalie Schneider[b], Peter Nick[b], and Alexander Rohrbach[a]

[a] Laboratory for Bio- and Nano-Photonics, Department of Microsystems Engineering, University of Freiburg, Georges-Koehler-Allee 102, 79110 Freiburg, Germany

[b] Molecular Cell Biology, Botanical Institute, Karlsruhe Institute of Technology, Kaiserstr. 2, 76131 Karlsruhe, Germany

[1] Present address: Lewis-Sigler Institute for Integrative Genomics, Princeton University, Washington Rd, Princeton, NJ 08544, USA

**SI Methods and Material**

The elastic relaxation forces along the filament $\int dF_{\kappa,MT}$ are reduced by filament friction $\int dF_{\gamma,MT}$ and act in lateral y direction. Due to the homolog constraint of the filament of constant length L and its connection to the optically trapped beads, the resulting elastic forces of the microtubule $F_{\kappa,MT}$ push the beads outwards in x-direction and are counteracted by the optical forces $F_{opt}$. Whereas the friction force $F_{\gamma,B}$ on the bead in the sensor trap (blue) is negligible small, the viscous force on the oscillating actor bead (red) counteracts the driving force $F_{drive}$.

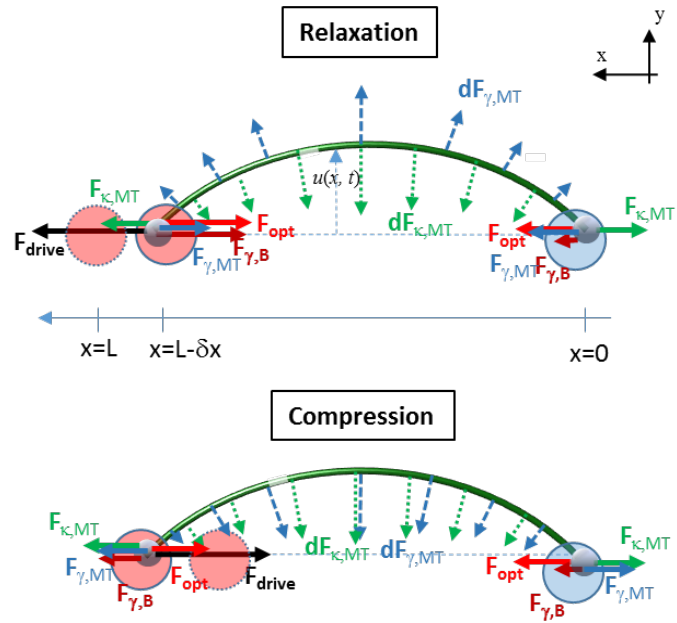

Fig. S1. Force diagrams of a buckling filament held by two optical traps in the case of relaxation or compression, for a pulling or a pushing driving force, respectively. The local elastic forces on different parts of the filament and the resulting elastic forces are shown in green, all viscous forces are shown in blue.

The tension free equation of motion for relaxation ($F_{drive}$ in positive x-direction) reads

$$-F_{opt1}(x_{B1}) - F_{\gamma_{B1}}(x_{B1}) + \tfrac{1}{2}F_{x,\kappa_{MT}}(u(x)) - \tfrac{1}{2}F_{x,\gamma_{MT}}(u(x)) = -F_{drive}(x_{B1}) \quad \text{for the left bead and}$$

$$+F_{opt2}(x_{B2}) + \underbrace{F_{\gamma_{B2}}(x_{B2})}_{\to 0} - \tfrac{1}{2}F_{x,\kappa_{MT}}(u(x)) + \tfrac{1}{2}F_{x,\gamma_{MT}}(u(x)) = 0 \quad \text{for the right bead.}$$

**Micro-rheology analysis**

The measured, frequency dependent displacements $x_{Bi}$, $y_{Bi}$ of bead $i$ as a response to an applied actuation force $F_x^{(j)}$, $F_y^{(j)}$ on bead $j$ is given by the response functions $A_x^{(i,j)}$, $A_y^{(i,j)}$ according to Eq. (S1)

$$x_{Bi}(\omega) \cdot \frac{1}{A_x^{(i,j)}(\omega)} = F_x^{(j)}(\omega)$$
$$y_{Bi}(\omega) \cdot \frac{1}{A_y^{(i,j)}(\omega)} = F_y^{(j)}(\omega)$$
(S1)

In active micro-rheology as it is used here, the driving force $F = \kappa \cdot x_L(t)$ is generate by a sinusoidal oscillation $x_L(t) = A_a \sin(\omega_a t)$ of one optical trap with stiffness $\kappa$, amplitude $A_a$ and driving frequency $\omega_a$. To obtain the complete spectrum $A(\omega)$, the experiment has to be repeated several times for different actuation frequency $\omega_a$ and evaluated according to Eq. (S1) for each frequency $\omega_a$. As explained in the main paper, the measured bead displacements $x_B$, $y_B$ are a superposition of the elastic trapping force, the viscous drag of beads and the wanted viscoelastic properties of the material under investigation, i.e., of the microtubule filaments in our case. Hence, the response function $A$ is a superposition of these contributions as well. As explained in (1, 2) given by Eq. (S2), this can be separated to obtain the pure viscoelastic response function $G_{MT}$ of the filament:

$$G_x^{(i,j)} = \frac{1}{4\pi L \cdot A_x^{(i,j)}} \left(1 - \kappa_{xi} A_x^{(i)} - \kappa_{xj} A_x^{(j)} + \kappa_{xi}\kappa_{xj}\left(A_x^{(i)} A_x^{(j)} - \left(A_x^{(i,j)}\right)^2\right)\right)$$
$$G_y^{(i,j)} = \frac{1}{8\pi L \cdot A_y^{(i,j)}} \left(1 - \kappa_{yi} A_y^{(i)} - \kappa_{yj} A_y^{(j)} + \kappa_{yi}\kappa_{yj}\left(A_y^{(i)} A_y^{(j)} - \left(A_y^{(i,j)}\right)^2\right)\right)$$
(S2)

Here, $\kappa^{(i)}$ and $\kappa^{(j)}$ are the trap stiffnesses of the corresponding traps which have to be determined independently by calibration (3, 4). Different pre-factors $4\pi L$ and $8\pi L$ for different directions take care of the hydrodynamic coupling $\gamma_{HC,y} = 2\gamma_{HC,x} = 8\pi L \eta \omega$ of distant sites (1) separated by the distance $L$. In the main paper, we usually used the notion parallel (∥) and perpendicular (⊥) instead of $x$ and $y$ according to the direction of oscillation with respect to the filament orientation.

To ensure correct results we tested the software implementation and measurement procedure for simple beads in water, where the viscoelastic response is known theoretically and measured experimentally (1). Since the motivation of the paper is to study the transport of mechanical stimuli, only the elastic components $G'^{(i,j)}$ (real part of $G$) are shown and discussed in the main paper. Viscous components (imaginary part of *G*) are shown in the SI Results (see below).

**The molecular architecture of differently polymerized and stabilized filaments**

We choose filaments polymerized in the presence of a non or slowly hydrolysable GTP analog (GMPCPP) in addition to filaments assembled with GTP since both filament types have significantly different mechanical properties (5) ultimately governed by different molecular configurations as illustrated in Fig. S2. After polymerization, GTP molecules in the microtubule lattice hydrolyze stochastically to GDP. While GTP and GTP-analog tubulins adopt a straight conformation, the hydrolysis at the β-tubulin leads to a kink of the GDP tubulin dimer resulting in an intrinsic strain in the microtubule lattice (6, 7). This conformational change is slowed down by Taxol (8) which binds on the inside of the hollow tube (9, 10) and has been used to theoretically recapitulate the tip structure and rates of assembly/disassembly of microtubules (11), the occurrence of long-lived arcs and rings in kinesin-driven gliding assays (12) and to transform MTs into inverted tubules facing their inside out by a specifically induced conformational change using spermine, a polyamine present in eukaryotic cells (13). Further, microtubules polymerized in the presence of slowly or non-hydrolyzable GTP analogs such as GMPCPP or γ-S-GTP have additional lateral inter-protofilament contacts between β-tubulins compared to GTP/GDP microtubules (5, 14). Assuming that the connection between individual αβ- tubulin dimers can be approximated by damped harmonic springs (15, 16), the damping of the intermolecular connections should affect the temporal response upon exertion of mechanical stimuli and thereby the transition frequency $\omega_t$.

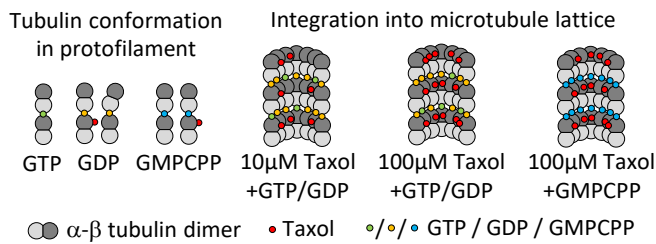

Fig. S2. Effect of GTP, GDP, GMPCPP, Taxol and combinations thereof, on molecular conformation of the αβ-tubulin dimer and corresponding binding sites.

# SI Results

## Lateral forces are negligible

In addition to the data shown in Fig. 2 of the main paper, we here compare the bead displacements along the x and y direction during a single filament rheology experiment at two oscillation frequencies $f = 0.1$Hz and $f = 100$Hz. As Fig. S 3 shows, the total contributions in lateral y-direction are negligibly small. Here, the lateral elastic MT buckling force $F_{\kappa MT,y}(x, x_{Bj})$ is increased (reduced) by the MT drag force $F_{\gamma MT,y}(x, x_{Bj})$ for deformation (relaxation). Both MT forces are equilibrated by the strong optical forces $F_{opt,y}(x_{Bj})$ and the weak viscous drag forces of the beads $F_{\gamma,y}(x_{Bj})$ in lateral direction. The sum of these forces is zero for all oscillation frequencies and phasings, i.e., $F_{opt,y}(x_{Bj}) + F_{\gamma,y}(x_{Bj}) + F_{\kappa MT,y}(x, x_{Bj}) + F_{\gamma MT,y}(x, x_{Bj}) \approx 0$. This situation is revealed in Fig. S 3:

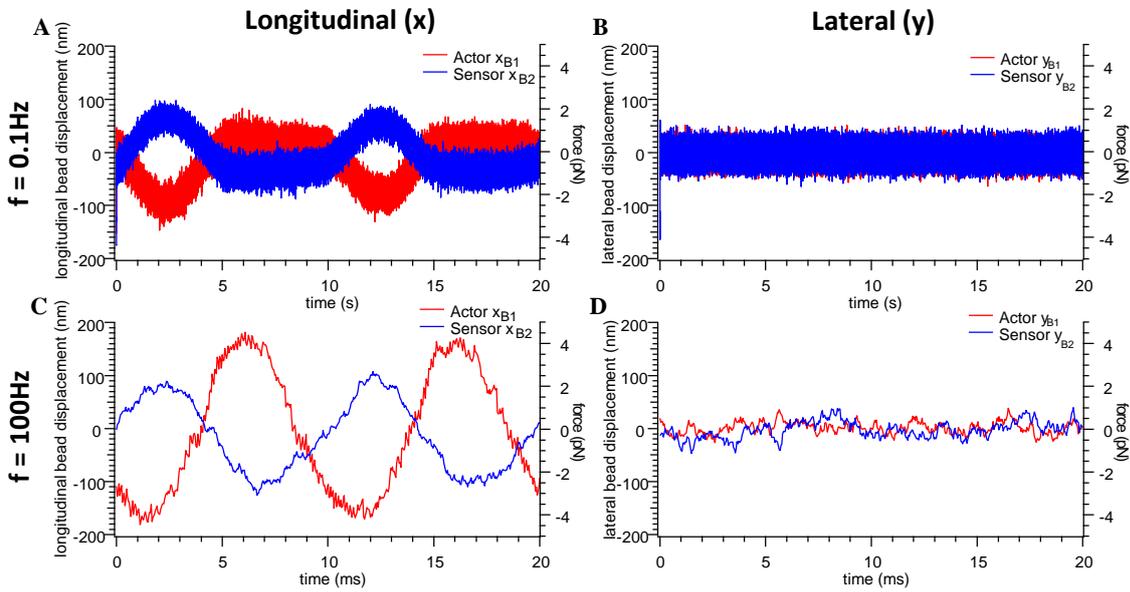

Fig. S 3. Responses of the actor and sensor bead resulting from the oscillatory driving of the actor bead in longitudinal (x) direction. Displacements and forces are large in longitudinal (x) direction (A and C), but very small in lateral (y) direction (B and D) for longitudinal oscillation frequencies at $f = 0.1$Hz (A and B) and $f = 100$Hz (C and D).

## Frequency dependent bead displacements

The displacements $x_{Bi}$ of the beads are governed by the elastic optical trapping force, the viscous drag force of the beads as well as the viscoelastic force from the microtubule filament according to Eq. (2) of the main article. In Fig. S4, we show the frequency dependence of the maximum actor and sensor bead displacement $|x_{Bi} - x_{Li}|$ during filament buckling and filament stretching. While an increase of the maximum amplitude of bead displacements of approximately one order of magnitude can be observed during buckling, bead displacements stay approximately constant and are proportional to the actor amplitude $A_a$ during filament stretching. Already here, the connection between the constant low frequency plateau of $G'$ and its power law rise above $f_t \approx 2$Hz to filament buckling can be anticipated. The viscoelastic contribution of the trapped beads alone moving in the purely viscous buffer medium is much smaller than the effect observed here and is dominated by the corner frequency $\omega_c = \kappa / 6\pi R_B \eta \approx 2500$ Hz of the position power spectral density $|x_B(\omega)|^2$ of the bead motion, which is much larger than the transition frequencies estimated for our MT constructs.

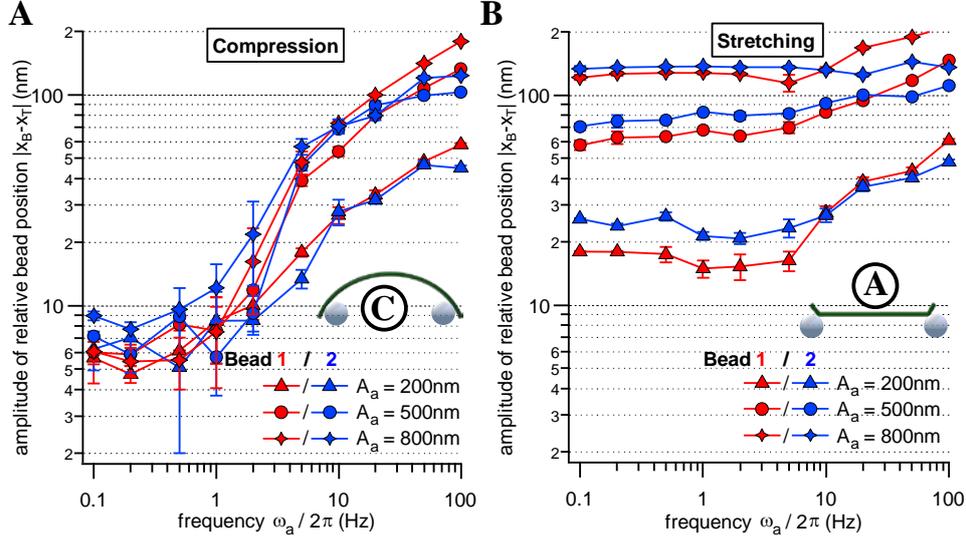

Fig. S4. Maximum displacement amplitudes $|x_{Bi} - x_{Li}|$ of both beads with index $i$ relative to their trap centers $x_{Li}$ during single filament oscillation. (A) During the compression (buckling) half period. (B) During the stretching half period.

**Estimation of the total viscous force**

To distinguish the absolute role of the different forces introduced in Eqs. (1) and (2) of the main paper, we determine these contributions in the following. The results shown here are the basis of Fig. 2D of the main paper.

The most obvious force is the viscous drag $F_{\gamma_B, \text{tran}}$ of bead translation. The actor bead approximately follows the sinusoidal movement of its trapping focus and is much larger than that of the sensor bead, which is neglected for this reason. The movement of the actor trap is $x_{L1}(t) = A_a \sin(2\pi f_a t)$ resulting in the velocity $v_{L1}(t) = \partial/\partial t\, x_{L1}(t) = 2\pi f_a A_a \cos(2\pi f_a t)$. Only the maximal force components are considered in the following. Hence, the translational viscous drag force is given by Eq. (S3) and shown by the yellow line in Fig. S5B.

$$F_{\gamma_B, tran}(f_a) = \gamma_B \max(v_{L1}) = 12\pi^2 R_B \eta A_a f_a \tag{S3}$$

The buckling filament causes both beads to rotate resulting in a rotational drag force $F_{\gamma_B, \text{rot}}$ governed by a varying angle $\varphi_A(\delta_L) = \sqrt{\frac{4\delta_L}{L_{MT}}}$ for different compressions $\delta_L$ of the filament as illustrated in Fig. S5A. Beads are rotated at the angular velocity $\omega_{B,rot} = \Delta\varphi_A f_a$ with $\Delta\varphi_A = \varphi_A(A_a)$ resulting in the viscous torque $M = \gamma_{B,rot}\, \omega_{B,rot}$ with $\gamma_{B,rot} = 8\pi R_B^3 \eta$ (17). This torque is balanced by a viscous force according to $M = R_B F_{\gamma_B,\text{tran}}$ resulting in Eq. (S4) and illustrated by green line in Fig. S5B.

$$F_{\gamma_B,\text{rot}}(f_a) = \frac{\gamma_{B,\text{rot}} \omega_{B,\text{rot}}}{R_B} = 8\pi\eta R_B^2 \sqrt{\frac{4A_a}{L_{MT}}} f_a \tag{S4}$$

To calculate the viscous drags $F_{\gamma_{MT},\parallel}(v_{\delta_L})$ and $F_{\gamma_{MT},\perp}(v_{\delta_L})$ acting on the MT during a parallel and perpendicular movement of the buckled filament with respect to the filament axis, the viscous drag coefficients of a rod $c_\parallel = \frac{2\pi\eta}{\ln(\frac{L}{D})-0.2}$ and $c_\perp = \frac{4\pi\eta}{\ln(\frac{L}{D})+0.84}$, the velocity of filament compression $v_{\delta L} = 2\pi f_a A_a$ and the buckling amplitude $u(\delta_L) = \sqrt{\left(\frac{L_{MT}}{\pi}\right)^2 - \left(\frac{L_{MT}-\delta_L}{\pi}\right)^2}$ as a function of filament compression $\delta_L$ is needed. The latter is deduced further below (see Eq. (S7)).

Assuming that the right side of the filament is approximately stationary while the left side moves at velocity $v_{\delta L}$, the parallel force component can be estimated by integrating the force per unit length of a small portion of the rod at position $x$ and moving at a velocity $\frac{x}{L_{MT}-\delta_L} v_{\delta_L}$ over its entire length as illustrated in Fig. S5A. The result is shown in Eq. (S5) and Fig. S5B by the blue line.

$$F_{\gamma_{MT},\parallel}(f_a) = c_\parallel v_{\delta_L} \int_0^{L-\delta_L} \frac{x'}{L_{MT}-\delta_L} dx' = \frac{2\pi^2 (L_{MT}-A_a)\eta}{\ln\left(\frac{L_{MT}}{D}\right)-0.2} A_a f_a \tag{S5}$$

In a similar fashion, the lateral force component can obtained by considering that both ends of the filament do not move in this direction in contrast to the filament center, and that the velocity of a small portion of the buckling rod at position $x$ is determined by the temporal change $\frac{du(\delta_L)}{dt} \sin\left(\frac{\pi x}{L_{MT}-\delta_L}\right)$ of the buckling amplitude. This results in Eq. (S6) and is described by the black line in Fig. S5B.

$$F_{\gamma_{MT},\perp}(f_a) = c_\perp \frac{d}{dt} u(\delta_L) \int_0^{L-\delta_L} \sin\left(\frac{\pi x'}{L_{MT}-\delta_L}\right) dx' = \frac{16(L_{MT}-A_a)^2}{\sqrt{L_{MT}^2-(L_{MT}-A_a)^2}} \frac{\eta A_a f_a}{\ln\left(\frac{L_{MT}}{D}\right)+0.84} \tag{S6}$$

For comparison, we also plotted the experimentally obtained frequency dependence of the total force of a bead alone (yellow markers) and of a bead / MT construct (red markers) together with the total force of all contributions estimated above (red line) in Fig. S5B for an oscillation amplitude $A_a = 600$nm. For the situation of a bead alone, the mean force $F \approx 0.5 pN \approx \kappa \langle |x_B| \rangle$ is constant until it intersects and overlaps with the translational viscous force $F_{\gamma_B,\text{tran}}$ of the bead, which is dominant at high frequency.

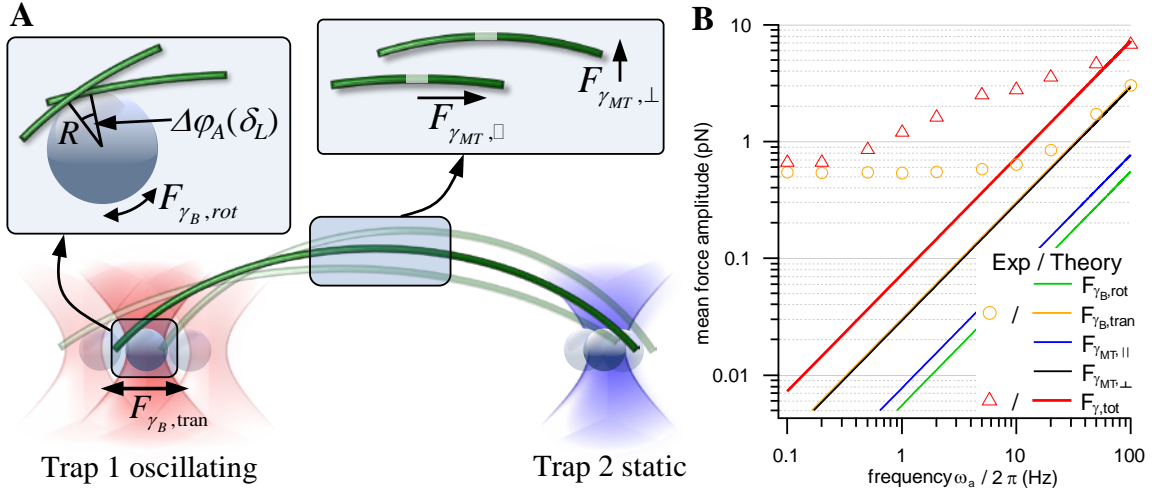

Fig. S5. Estimation of viscous forces. (A) Illustration of experimental situation for a single filament held by two optically trapped beads, one static (trap 2) and one oscillating (trap 1). (B) Frequency dependence of theoretically obtained and experimentally measured force contributions.

For the experimental situation with MT filament, the total force increases continuously but still intersects with the sum of all estimates of the viscous contributions made above at approximately $f = 100$Hz. This indicates a strong additional contribution from the MT, very likely of elastic nature as we obtained by the micro-rheology analysis for $G'$.

**Contributions of deformation modes to $G'(\omega)$**

As stated in the main paper, we estimate to excite $N = 3$ deformation modes at oscillation frequencies up to $f_a = 100$Hz. The contributions of each additional mode are illustrated in Fig. S6 where we plotted the theoretical slope of $G'(\omega, N) = \frac{1}{2.16\pi^2} q_1^4 k_B T \ell_p \, \text{Re}\left(\sum_{n=1}^{N} \frac{1}{n^4 + i\omega/\omega_1}\right)^{-1}$ for the sum of different deformation modes $N = 1, 2, 3$ and $10$ (solid lines) according to Eq. 5. We normalized the shear modulus $G'(\omega, N)/G'(0, N)$ to obtain a better estimate for the relative contributions. We also plotted the difference $G'(\omega, N) - G'(\omega, N-1)$ (dashed lines) to indicate the influence of a single mode and at which frequency the next mode kicks in.

The first mode, causing a constant plateau, is dominant for low frequencies up to $\omega \approx \omega_1$. Higher deformation modes are excited and result in an increased filament stiffness for frequencies $\omega \geq 3\omega_1 = \omega_{t2} = \omega_t$ ($N = 2$) and $\omega \geq 20\omega_1 = \omega_{t3}$ ($N = 3$). Modes higher than $N \geq 4$ are relevant only for frequencies $\omega > 100\omega_1$, which is beyond the experimentally addressed frequency limit in our study.

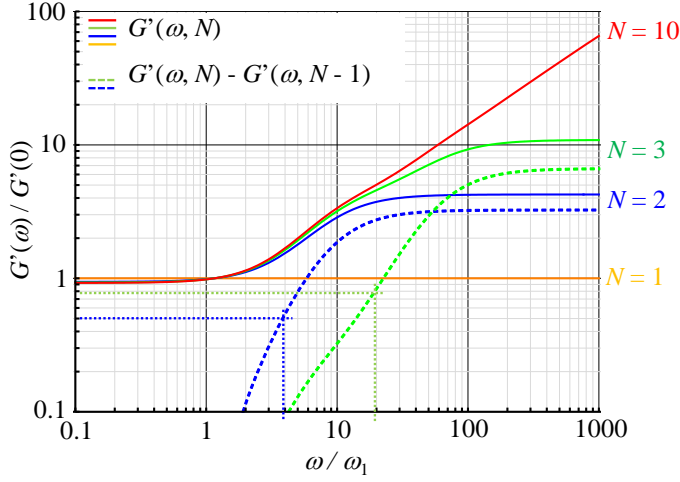

Fig. S6. Theoretical estimate of mode dependence of $G'(\omega)$. The elastic modulus $G'(\omega, N)$ is shown including a different number $N$ of deformation modes $n$ (solid lines) as well as the difference of $G'(\omega, N) - G'(\omega, N-1)$ for different modes (dashed lines).

**Viscous components of single filaments**

The viscous modulus $G''(\omega)$ of a simple bead in a ideally viscous solution such as water is $G''(\omega) = \gamma_B \cdot \omega / 6\pi R_B = \eta\omega$ (1). Similarly, the viscous component of a rod can be expected to be $G''(\omega) = \gamma_{MT} \cdot \omega / 4\pi L \sim \eta\omega$, i.e., linear with respect to the frequency $\omega$. Indeed, this is what we observe for the viscous modulus of all single filaments as shown in Fig. S7A. The results still depend on the length $L_{MT}$ of the filament due to the logarithmic dependence of $\gamma_{MT} \sim 1 / (\ln(L_{MT} / D) + 0.84)$ on $L_{MT}$ but does not depend on filament stabilization, contrary to $G'(\omega)$ as described in the main paper.

A comparison to the theoretical prediction shown in Fig. S7B reveals that high deformation modes $n \geq 2$ only slightly change this linear relationship and only for relatively high frequencies $\omega > 10\omega_1$, which is roughly the maximum frequency we resolve in our experiments ($\omega_1 \approx 4$Hz typically, see main paper).

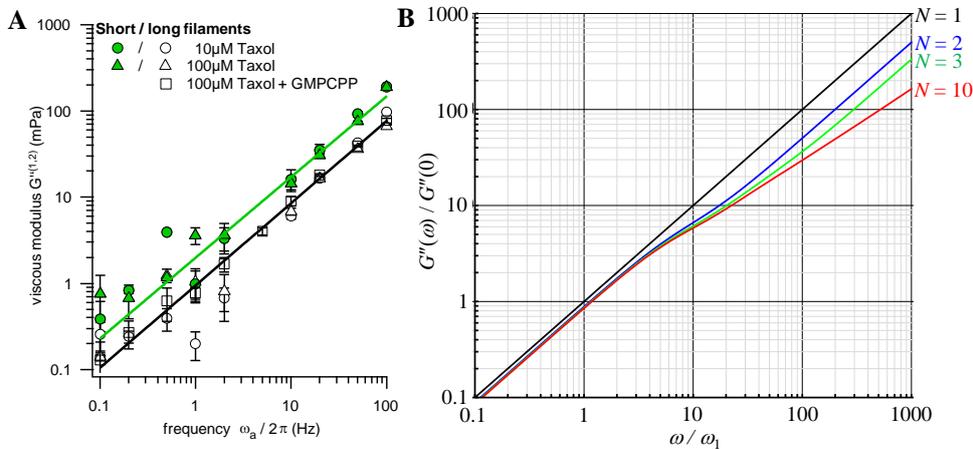

Fig. S7. Viscous components $G''(\omega)$ of single filaments. (A) Experimental results. Solid lines represent linear fits to the data. (B) Theoretical slopes for different number of oscillation modes $N$.

**Viscous components of a linear connection of MTs**

Similarly to the viscous components of single MT filaments, the viscous components $G''^{(1,2)}_{\perp,\parallel}$ and $G''^{(1,3)}_{\perp,\parallel}$ of a linear connection of filaments are again linear with respect to frequency $\omega$ as shown in Fig. S8 by power law fits with free exponent $p \approx 1$ in all cases.

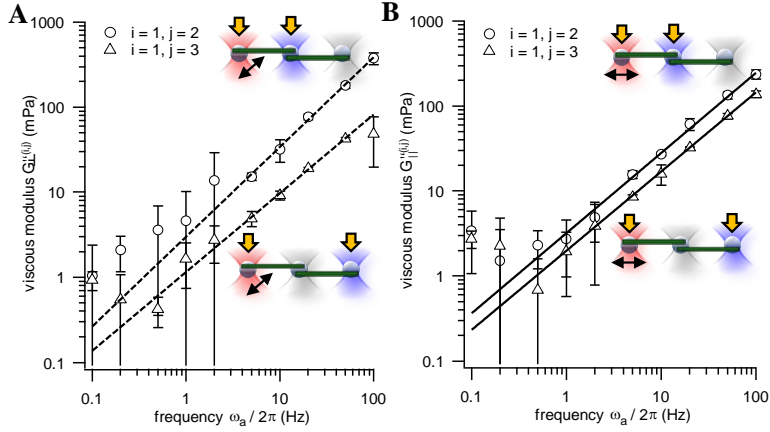

Fig. S8. Viscous components $G''(\omega)$ of a linear connection of filaments. Solid and dashed lines indicate approximately linear fits to the data.

**Pre-stress in triangular networks**

Free floating filaments are subject to Brownian forces and sometimes bend heavily. This can cause pre-stress during the construction of a network, i.e., the subsequent attachment of a filament to optically trapped beads. This effect becomes more prominent if the number of filaments of a network increases. Fig. S9 shows the elastic modulus $G'$ of three different equilateral triangles with a side length of 15µm. For both oscillation directions, the plateau of $G'$ is larger for the connection of the first filament 1→2 and smaller for the connection of the second filament 1→3 for the first two triangles, compared to the third triangle where the elastic moduli for both connections are approximately equal. This clearly indicates a pre-stress of the first MT compared to the second filament.

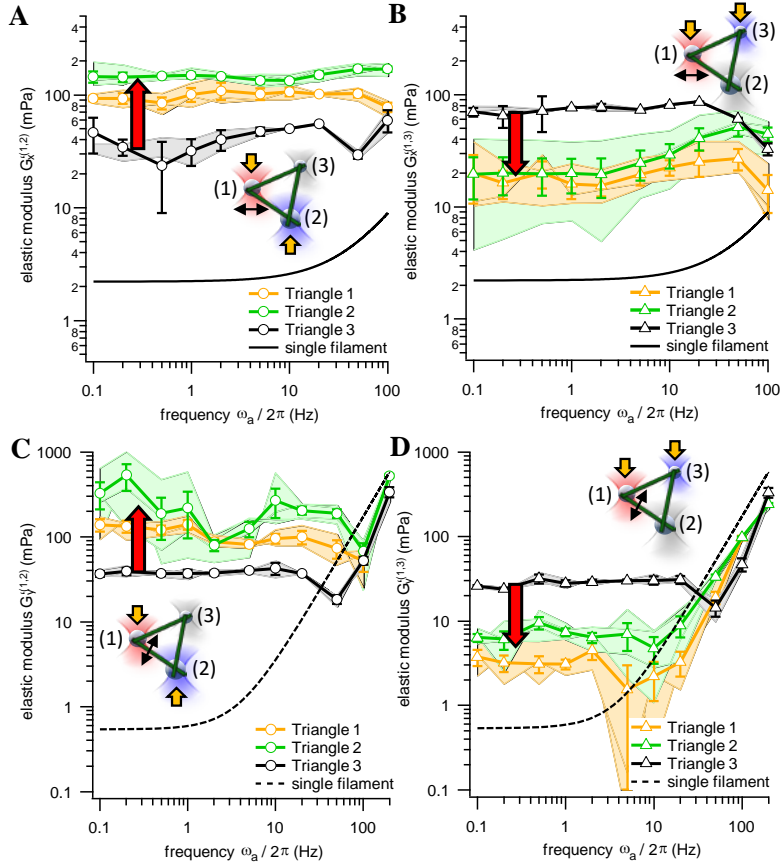

Fig. S9. Pre-stress in an equilateral triangle. (A) and (B) $G'(\omega)$ for a radial oscillation along $x$ for the connection 1→2 (A) and 1→3 (B). (C) and (D) $G'(\omega)$ for a tangential oscillation along $y$ for the connection 1→2 (C) and 1→3 (D)

**Viscous components of triangular networks**

Again, we observe a linear relation between the viscous components $G''(\omega)$ of filaments in a triangular networks and the frequency $\omega$ as shown in Fig. S10 together with power law fits with free exponent $p \approx 1$. For the tangential oscillation along $y$, the viscous component $G''^{(1,2)}$ of the first filament deviates strongly from the expected linear response for the first two triangle constructs, indicating that maybe the connection to either of the beads (1) or (2) was not perfect.

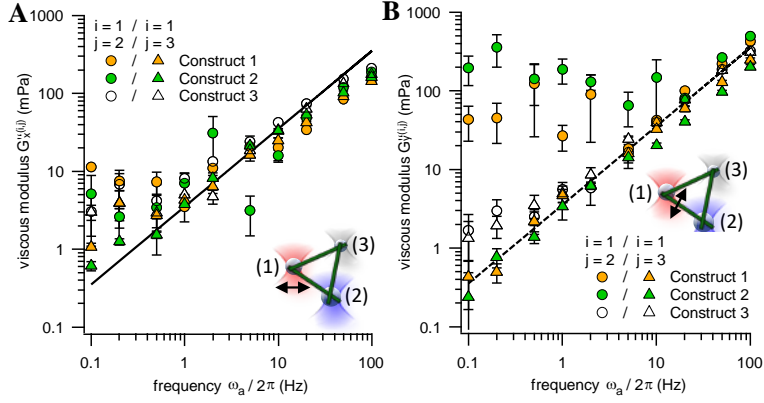

Fig. S10. Viscous components $G''(\omega)$ of a triangular connection of filaments for actor bead displacement along x (left) and along y (right). Solid and dashed lines indicate approximately linear fits to the data.

## Comparison of transition frequencies

As described in the main paper, we found that he transition frequency $\omega_t$, separating the constant plateau value of $G'$ for low frequencies from the high frequency rise approximately proportional to $\omega^{1.25}$, depends on filament length, stabilization, polymerization and especially on the geometry of the network. This is summarized in Fig. S11A. The transition frequency is the highest for the triangular network, which is also the stiffest. The difference of filament stabilization is clearly visible for long filaments. There is also a clear difference visible for different oscillation directions of all geometries, where the transition frequency is much smaller for an oscillation lateral to the filament axis, indicating much faster stiffening in this direction.

In order to analyze how well the experimental transition frequencies match the theoretical predictions according to Eq. 3 of the main paper, we plotted the transition frequency as a function of MT contour length as shown in Fig. S11B. Here, we included the length dependence of the MT persistence length $l_p(L, \omega = 0) = l_p^\infty / \left(1 + (l_c / L)^2\right)$ according to Pampaloni (18). For the persistence length $l_p^\infty$ of MTs much longer than a critical length $l_c = 21\mu m$, we used the values for $L = 15\mu m$ long MTs obtained in this study to reflect the different stabilizations.

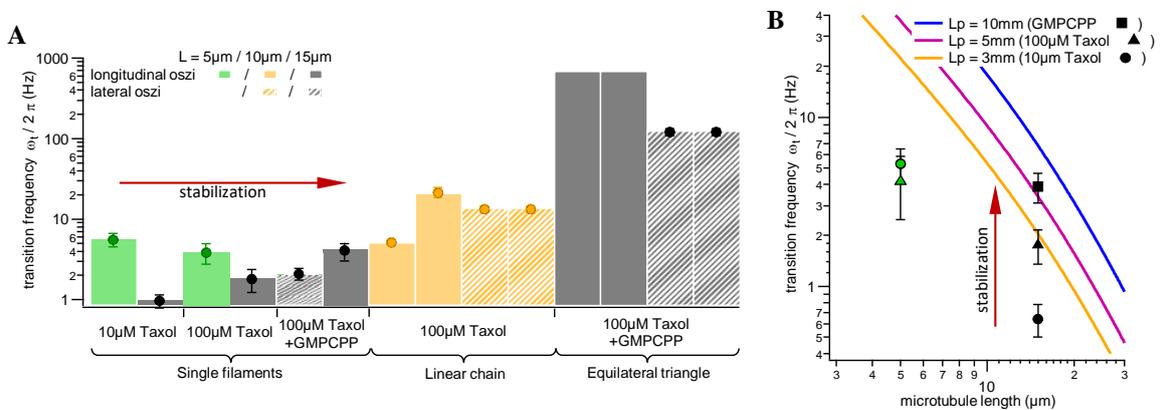

Fig. S11. Transition frequencies for different network geometries, filament stabilizations and polymerizations.

**Geometric effects of beads**

MT filaments are attached laterally to the anchor beads with a diameter $d = 1062$nm. This results in a torque on the beads during filament buckling because the optical trapping force acts on their geometric centers. This causes the point of attachment of the filament to a bead to rotate. Hence, the precisely measured distance $\Delta x_L + x_{B1} - x_{B2}$ between both beads does not coincide with the actual projected length $p_{MT}$ of the buckled filament as illustrated in Fig. S12A. For convenience, we assume the symmetric case with $x_{B1} = x_{B2} = x_B$ and $\Delta_1 = \Delta_2 = \Delta_B = R_B \cdot \sin(\varphi) \approx R_B \cdot \varphi$ in the following. Neither the actual compression $\delta_L$ given by Eq. (S7) nor the rotation angle $\varphi^2 = 4\delta_L / L_{MT}$ can be measured directly.

$$\delta_L = L_{MT} - p_{MT} = L_{MT} - (\Delta x_L + 2x_B - 2\Delta_B) \quad (S7)$$

However, both unknowns depend on each other leading to the quadratic relation $\varphi^2 + \frac{8R_B}{L_{MT}}\varphi + 4\left(\frac{\Delta x_L + 2x_B}{L_{MT}} - 1\right) = 0$ and ultimately to an expression for the angle $\varphi$ which only depends on known or measured quantities given by Eq. (S8).

$$\varphi = \frac{4R_B}{L_{MT}}\left(\sqrt{1 + \frac{L_{MT}}{4R_B^2}(L_{MT} - \Delta x_L - 2x_B)} - 1\right) \quad (S8)$$

This can be substituted in Eq. (S7) to calculate the actual compression $\delta_L$ and to plot force compression curves, i.e., the buckling force $F = \kappa_1 \cdot x_{B1} + \kappa_2 \cdot x_{B2}$ versus the compression $\delta_L$ as we show for two filaments with different length in Fig. S12C+D. The data shown here were obtained in a quasi-equilibrium where we moved trap 1 in discrete steps of $\Delta x_{L1} = 50$nm every $\Delta_t = 100$ms.

In the ideal case, i.e., a perfectly axial application of force on the filament, the MT should not buckle, i.e., $\delta_L(F < F_{crit}) = 0$, until a finite critical buckling force $F_{crit} = \pi^2 EI / L^2_{MT}$ is reached, above which the filament behaves like a spring with spring constant $\kappa_{MT}$, i.e., $\delta_L(F > F_{crit}) = (F - F_{crit}) / \kappa_{MT}$ (17). Here, we observe a nearly exponential dependence of the force on the compression for small $\delta_L < 400$nm and a linear dependence for $\delta_L > 400$nm. This is due to the imperfect, lateral application of the force on the filament. We indicated this behavior in Fig. S12C+D by the blue fits with fit function $F(x) = F_{crit}(1 - e^{-\frac{x}{x_0}}) + \kappa_{MT}x$. From this, we determined the critical force $F_{crit}(L_{MT})$ for various filaments of different length as shown in Fig. S12B. As expected from the ideal case, the critical force increases with decreasing filament length, however, not as fast as the ideal $1 / L^2_{MT}$ dependence predicts. This is due to the length dependence of the persistence length $l_p(L_{MT}) = l_p^\infty(1 + \frac{L_c^2}{L_{MT}^2})^{-1}$ (18), which has been used together with $F_{crit}(L_{MT}) = \pi^2 l_p(L_{MT}) k_B T / L^2_{MT}$ to fit the data. $L_c = 3.3$µm is the critical filament length above which the persistence length levels to a plateau $l_p^\infty = 2.2$mm in our case. Pampaloni et al. (18) obtained $L_c = 21$µm and $l_p^\infty = 6$mm from their thermal fluctuation data. This deviation is likely a result of different filament polymerization and stabilization. However, we did not consider this geometric effect in our rheology experiments. We expect this to have only a minor effect on the measured viscoelastic properties $G_{MT}$ of the filaments, but this has to be tested and included into the theory in the future.

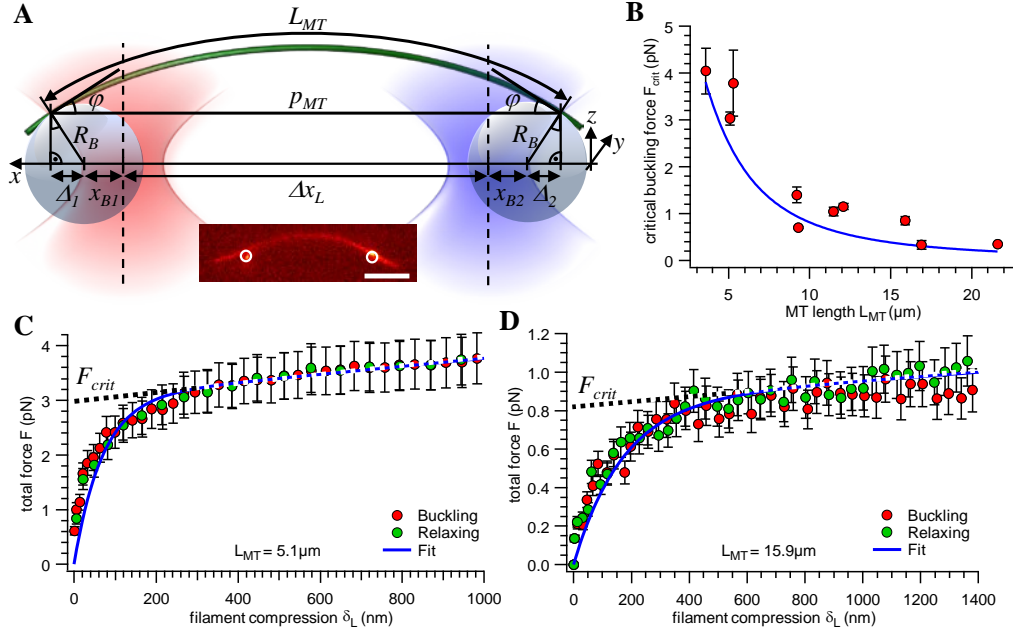

Fig. S12. Geometric effect of filament attachment to the spatially extended anchor beads. (A) Sketch of the experimental situation. (B) Dependence of the critical force $F_{crit}(L_{MT})$ on the filament length $L_{MT}$. (C) + (D) Plot of the total force $F(\delta_L)$ as a function of filament compression $\delta_L = L_{MT} - p_{MT}$ for a short MT ($L_{MT} = 5.1\mu m$, Fig. C) and long ($L_{MT} = 15.9\mu m$, Fig. D) MT. Experiments have always been repeated for an increasing buckle and back relaxation to the straight equilibrium position.

**Microtubule buckling amplitude**

For the integration step in equation (5) of the main manuscript, we need to derive an expression for the buckling amplitude $u_{qn}(\delta_L)$ as a function of the compression $\delta_L(t)$ of the filament with shape $u(x) = \sum_n u_{qn}(\delta_L)\sin(q_n x)$ and an arbitrary bending mode $q_n = \frac{n\pi}{L_{MT} - \delta_L}$.

Considering the invariant arc length $L' = \int \sqrt{1+(\frac{\partial u}{\partial x})^2}dx \stackrel{!}{=} L_{MT}$ of the buckled filament one obtains:

$$L' = \int_0^{L-\delta_L} \sqrt{1+\left(\sum q_n u_{qn} \cos(q_n x)\right)^2} dx \quad (S9)$$

Since this square root cannot be solved analytically, we investigate the buckling amplitude $u_{qn}$ as a function of arc length $L'_n$ for single bending modes n, where the ground mode n=1 allows to estimate the maximum possible deflection $u_{q1} > u(x)$.

Substituting $\varphi = q_n x$ leads to

$$L'_n = \int_0^{n\pi} \sqrt{\tfrac{1}{q_n^2} + u_{qn}^2 \cos^2(\varphi)}\, d\varphi = 2n \int_0^{\pi/2} \sqrt{\tfrac{1}{q_n^2} + u_{qn}^2 \cos^2(\varphi)}\, d\varphi \tag{S10}$$

Using $\cos^2(\varphi) = 1 - \sin^2(\varphi)$, equation (S10) can be rewritten in the form

$$L'_n = 2n \sqrt{\tfrac{1}{q_n^2} + u_{qn}^2} \int_0^{\pi/2} \sqrt{1 - \tfrac{u_{qn}^2}{\tfrac{1}{q_n^2} + u_{qn}^2} \sin^2(\varphi)}\, d\varphi \tag{S11}$$

Substituting $u'_{qn} = 2n\sqrt{\tfrac{1}{q_n^2} + u_{qn}^2}$ and $\phi = \dfrac{u_{qn}}{\sqrt{\tfrac{1}{q_n^2} + u_{qn}^2}} = 2n \dfrac{u_{qn}}{u'_{qn}}$ the relation

$$L_{MT} \overset{!}{=} L'_n = u'_{qn} E_{is\phi}(\phi(\delta_L)) \tag{S12}$$

can be deduced. $E_{is\phi}(\phi) = \int_0^{\pi/2} \sqrt{1 - \phi^2 \sin^2(x)}\, dx$ is the complete elliptical integral of the second kind, which cannot be solved analytically. However, an approximation formula $E_{is\phi} = \tfrac{\pi}{2}\left(1 - \tfrac{1}{4}\phi^2 - \tfrac{9}{192}\phi^4 + O(\phi^6)\right)$ can be used (19), leading to a first order approximation $L_{MT} = u'_{qn} \tfrac{\pi}{2} = n\pi \sqrt{\tfrac{1}{q_n^2} + u_{qn}^2}$ and ultimately to

$$u_{qn}(\delta_L) = \tfrac{1}{n\pi} \sqrt{L_{MT}^2 - (L_{MT} - \delta_L)^2} = \tfrac{1}{n\pi} \sqrt{2 L_{MT} \delta_L(t) - \delta_L^2(t)} \tag{S13}$$

As shown in Fig. S13 for a microtubule with length $L_{MT} = 10\,\mu m$, the buckling amplitude $u_{qn}$ increases rapidly with small compressions $\delta_L < 50$ nm and then approximately linearly for larger compressions $\delta_L > 100$ nm. The buckling amplitude decreases with the mode number n. However, the sum of all deformations in a filament is always smaller than the ground mode buckling, i.e. $u(x) < u_{q1}$.

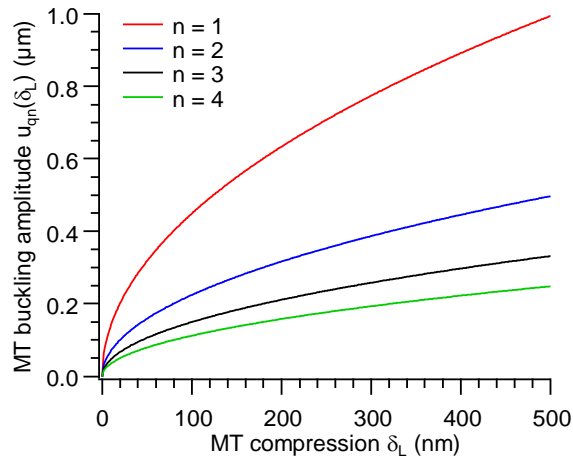

Fig. S13. Buckling amplitude $u_{qn}(\delta_L)$ as a function of filament compression $\delta_L$ and mode number $n$ for a filament with length $L_{MT} = 10\mu m$.

**Linear system theory**

Our theoretical description as well as our analysis assume a linear relationship between the microtubule buckling amplitude $u_{qn}$ and the driving force $F_D$, such that $\tilde{u}_{qn}(\omega) = \alpha_{qn}(\omega)\tilde{F}_D(\omega)$.

In order to test whether the buckling responses of single MTs are linear with force, we analyzed the force dependency $F(\delta_L)$ on a stepwise MT compression by $\delta_L$. Fig. S14 shows that $F(u_{qn})$ increases indeed roughly linearly for not too large buckling amplitudes $u_{qn}(\delta_L)$, in accordance with to Eq. S13.

Linearity: By analyzing the normalized $\chi^2$ value as a function of the number of data points included in a linear fit to the data, we find an approximately linear response up to $u_{qn} \leq 300$ nm for short microtubules (L = 5µm) and $u_{qn} \leq 1.4$µm for long filaments (L = 15µm). This is equivalent to an oscillation amplitude of the laser trap $x_L < 500$nm for short and $x_L < 1200$ nm for long microtubules, respectively. In our rheology experiments, we usually analyze the microtubule response for three different oscillation amplitudes $A_a = 200$ nm, $A_a = 400$ nm, and $A_a = 600$ nm. Hence, linear response is well fulfilled for long microtubules and at least for the two smaller oscillation amplitudes for short filaments. In addition, we always compare the results for different oscillation amplitudes to each other and never observe a significant difference.

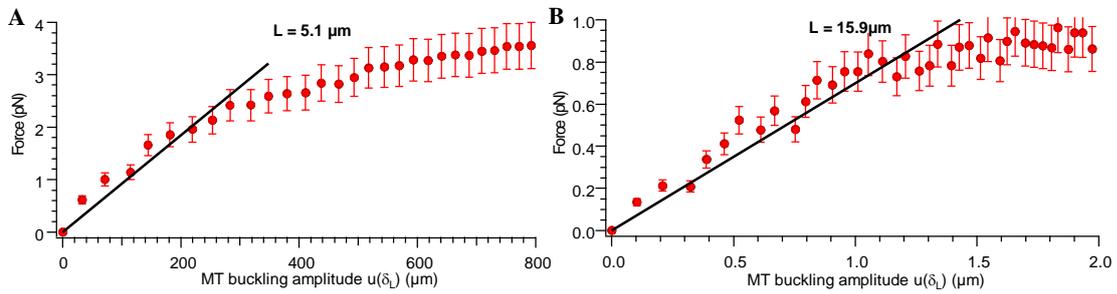

Fig. S 14. Compression force as a function of filament buckling amplitude for a short (A) and long (B) microtubule at $\omega \approx 0$.

**Local forces acting along the filament**

The curve of a filament deformed in the mode $q_n$ can be described by the vector $\vec{r}(x) = \begin{pmatrix} x \\ u_{qn} \sin(q_n x) \end{pmatrix}$ and has a tangent vector

$$\vec{T}(x) = \frac{d\vec{r}(x)}{dx} \bigg/ \left|\frac{d\vec{r}(x)}{dx}\right| = \frac{1}{\sqrt{1+u_{qn}^2 q_n^2 \cos^2(q_n x)}} \cdot \begin{pmatrix} 1 \\ u_{qn} q_n \cos(q_n x) \end{pmatrix}, \quad (S14)$$

such that the normal vector is $\vec{N}(x) = \begin{pmatrix} T_y(x) \\ -T_x(x) \end{pmatrix}$. From this, the local elastic force on an infinitesimal section of the filament can be calculated, according to

$\vec{f}(x) = EI \cdot \frac{d^4 u(x)}{dx^4} \vec{N}(x) = EI \cdot \frac{u_{qn} \cdot q_n^4 \cdot \sin(q_n x)}{\sqrt{1+u_{qn}^2 q_n^2 \cos^2(q_n x)}} \begin{pmatrix} u_{qn} q_n \cos(q_n x) \\ -1 \end{pmatrix}$, leading to a total force which acts only in y direction

$$\vec{F}_{\kappa,MT}(x) = \int_0^{L-\delta_L} \vec{f}(x)dx = \frac{EI \cdot q_n^3}{u_{qn}} \cdot \left[ \begin{pmatrix} -\sqrt{1+u_{qn}^2 q_n^2 \cos^2(q_n x)} \\ \operatorname{asinh}(u_{qn} q_n \cos(q_n x)) \end{pmatrix} \right]_0^{L-\delta_L} = \begin{pmatrix} 0 \\ F_y \end{pmatrix} \quad (S15)$$

We like to point out the strong dependence of the local bending force on the third power of the mode number n, meaning that the highest present order always dominates the buckling force.

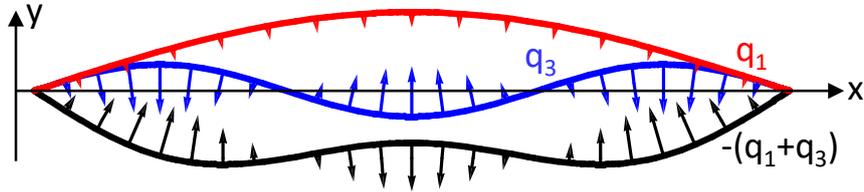

Fig. S 15. Filament deformation for the first and the third mode and the summation of both (flipped vertically). The resulting local force vectors are perpendicular to the filament tangent.

**Additional references**